\documentclass[aip,graphicx,reprint]{revtex4-1}

\usepackage{siunitx}
\usepackage{graphicx}
\usepackage{todonotes}
\usepackage{mathtools}
\usepackage{amssymb, amsmath}
\usepackage{multirow,tabularx}
\usepackage[T1]{fontenc}
\usepackage[english]{babel}

\usepackage[capitalise]{cleveref}

\DeclarePairedDelimiter\ket{\lvert}{\rangle}

\newcommand{\braket}[2]{\left\langle #1 | #2 \right\rangle}

\newcommand{\GaAs}{\ensuremath{GaAs}}
\newcommand{\AlAs}{\ensuremath{AlAs}}
\newcommand{\AlGaAs}{\ensuremath{Al_{0.15}Ga_{0.85}As}}
\newcommand{\nm}{\nano\metre}
\newcommand{\um}{\micro \metre}

\newcommand{\X}{\ensuremath{X}}
\newcommand{\XX}{\ensuremath{XX}}
\newcommand{\Xm}{\ensuremath{X^-}}
\newcommand{\Xp}{\ensuremath{X^+}}
\newcommand{\XXtoX}{\XX\mathord{-}\X}
\newcommand{\DetResP}[1]{\mathcal{N}(\delta_{#1})}
\newcommand{\DetResPP}[1]{\mathcal{N}(\delta_{#1})}
\newcommand{\FSS}{\ensuremath{\Delta_\text{FSS}}}
\newcommand{\Neg}{n}

\begin{document}


\title{An ultra-compact deterministic source of maximally entangled photon pairs}

\author{M. Langer}
    \affiliation{%
    Institute for Emerging Electronic Technologies, IFW Dresden, Helmholtzstraße 20, 01069 Dresden, Germany
    }

\author{P. Ruchka}
    \affiliation{
    4th Physics Institute and Research Center SCoPE, University of Stuttgart, 70569 Stuttgart, Germany
    }

\author{A. Rahimi}
    \affiliation{%
    Institute for Emerging Electronic Technologies, IFW Dresden, Helmholtzstraße 20, 01069 Dresden, Germany
    }

\author{S. Jakovljevic}
    \affiliation{ 
    4th Physics Institute and Research Center SCoPE, University of Stuttgart, 70569 Stuttgart, Germany
    }

\author{Y. G. Zena}
    \affiliation{%
    Institute for Emerging Electronic Technologies, IFW Dresden, Helmholtzstraße 20, 01069 Dresden, Germany
    }

\author{S. A. Dhurjati}
    \affiliation{%
    Institute for Emerging Electronic Technologies, IFW Dresden, Helmholtzstraße 20, 01069 Dresden, Germany
    }

\author{A. Danilov}
    \affiliation{%
    Institute for Emerging Electronic Technologies, IFW Dresden, Helmholtzstraße 20, 01069 Dresden, Germany
    }

\author{M. Pal}
    \affiliation{%
    Institute for Emerging Electronic Technologies, IFW Dresden, Helmholtzstraße 20, 01069 Dresden, Germany
    }

\author{R. Bassoli}
    \affiliation{
    Quantum Communication Networks research group, Deutsche Telekom Chair of Communication Networks, Technische Universität Dresden, Germany
    }

\author{F. H. P. Fitzek}
    \affiliation{
    Deutsche Telekom Chair of Communication Networks, Technische Universität Dresden, Germany
    }

\author{O. G. Schmidt}
    
    \affiliation{
    Research Center for Materials, Architectures and Integration of Nanomembranes (MAIN), Chemnitz University of Technology, Chemnitz, Germany
    }

\author{H. Giessen}
    \email{giessen@pi4.uni-stuttgart.de}
    \affiliation{
    4th Physics Institute and Research Center SCoPE, University of Stuttgart, 70569 Stuttgart, Germany
    }

\author{C. Hopfmann}
    \email{caspar\_arndt.hopfmann@tu-dresden.de}
    \affiliation{%
    Institute for Emerging Electronic Technologies, IFW Dresden, Helmholtzstraße 20, 01069 Dresden, Germany
    }
    \affiliation{
    Quantum Communication Networks research group, Deutsche Telekom Chair of Communication Networks, Technische Universität Dresden, Germany
    }

\date{\today}

\begin{abstract}

We present an ultra-compact source of maximally entangled on-demand photon pairs. Our results are based on coupling of single $\GaAs$ quantum dots that are embedded in monolithic micro-lenses to a single-mode fiber with directly attached to 3D-printed micro-optics ($NA=0.6$) inside a cryogenic environment. This approach, which is geared towards future integration into industrial environments, yields state-of-the-art entangled photon pair creation performance while retaining flexibility and adjustability required for long-term operation of such a device – all while dramatically reducing the overall system footprint. We demonstrate near diffraction-limited performance and hyperspectral imaging utilizing a 3D-printed micro-objective with a full width at half maximum resolution limit of \qty{604 \pm 16}{\nm} when operating the system at a cryogenic temperature of \qty{3.8}{\kelvin}.  Furthermore, we prove that this system can be used to achieve single photon emission rates of \qty{392 \pm 20}{\kilo\hertz} at a \qty{76}{\mega\hertz} pump rate and purities of \qty{99.2 \pm 0.5}{\percent} using two-photon resonant excitation. Utilizing the exciton-biexciton emission cascade available in $\GaAs$ quantum dots under resonant excitation, near maximally entangled photon pairs with peak entanglement negatives $2n$ of \num{0.96 \pm 0.02} in a \qty{4}{\pico \second} time window, and \num{0.81 \pm 0.01} when averaged over one exciton lifetime, are demonstrated.
\end{abstract}


\maketitle 

\section{\label{sec:Intro} Introduction}

In the quest for realizing quantum communication networks \cite{Kim2008, Wei2022}, significant progress has been made in the past decade by realizing high-performing quantum light sources \cite{Claudon2010, Dousse2010, Somaschi2016, Chen2018, Wang2019, Liu2019, Lu2021, Tomm2021, Shooter2020, Uppu2020, Hopfmann2021}. In order to realize multipartite quantum networks, entangled photon sources are a key component \cite{Loock2020, Schimpf2021}. While sources based on parametric down-conversion have been very successful \cite{Wang2016, McCutcheon2016} - partly due to their relative ease of implementation using non-linear optics - the performance of these sources remains fundamentally limited to their classical Poissonian statistics \cite{Zhang2021a}. Specifically, on-demand entangled photon pair sources with high efficiencies, entanglement fidelities and indistinguishabilities are crucial in order to enable Bell-state interferences necessary to build multipartite quantum information exchange networks and quantum repeater systems \cite{Kim2008, Loock2020} enabling applications such as distributed quantum computing \cite{Cacciapuoti2020} and physically secure communication \cite{Basset2021}. Quantum light sources such as semiconductor quantum dots (QDs), color centers in diamond and trapped atoms or ions can potentially realize this kind of quantum light sources as they do not suffer from the same limitations due to their Fock state-like quantum photon number distributions \cite{Stevens2013, Loock2020}. In order to attain highly coherent and efficient quantum light sources, all of these material systems require highly controlled environments - in solid-states systems this is commonly achieved by cooling to cryogenic temperatures, typically below \qty{10}{\kelvin}. Due to premise, it is highly challenging to miniaturize the realized sophisticated sources and make them suitable for industrial environments - such as server rooms - which is a prerequisite to realize practical implementations of quantum networks on a large scale. Furthermore, as today's communication infrastructure relies heavily on data exchange via optical fibers and unstabilized free-space optical systems are unsuitable for industrial environments - such as $19 \, in$-rack systems - compact, mobile and in situ fiber-coupled quantum light sources are very desirable.
While deterministically fabricated, compact and fiber-coupled single photon sources based on $InGaAs$ QDs have been explored \cite{Schlehahn2018, Musial2020, Bremer2020}, their performance is orders of magnitude lower compared to cavity-based light extraction \cite{Somaschi2016, Wang2019,Tomm2021} and free-space collection approaches \cite{Claudon2010, Chen2018}. 
To address this challenge, a viable design of an efficient, ultra-compact and fiber-coupled entangled photon pair source based on the droplet-etched $\GaAs$ QD platform is presented in this work. By realizing such a device, we expect our work to contribute towards scalable out-of-lab realizations of future entanglement-based quantum communication networks.

\section{\label{sec:Exp_Setup} System Design}

\begin{figure}
\includegraphics[width=0.5\textwidth]{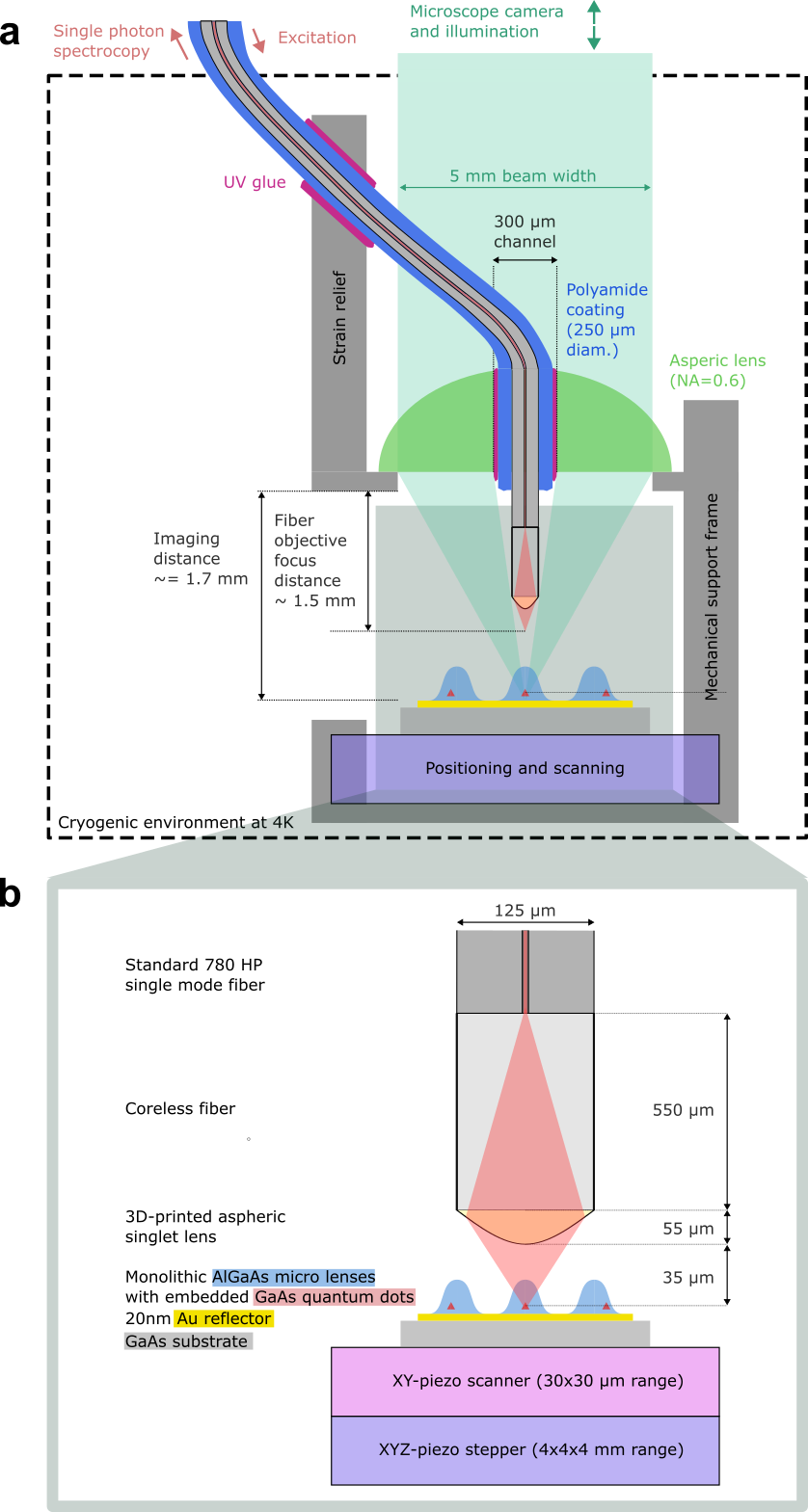}%
\caption{\label{fig:Setup_SNOM} Schematic illustration of the design of the ultra-compact fiber-coupled entangled photon pair source. The entangled photons emitted by the $\GaAs$ quantum dots embedded in a monolithic microlens are collected directly in the cryostat in a single mode fiber using a 3D-printed micro-objective. a) Overall system optical and mechanical design overview, including supplementary free-space microscopy for rough positioning of fiber micro-objective, fiber strain relief, and mechanical support. b) Detailed illustration of design of the collection optics and spatial positioning of entangled photon pairs from single $GaAs$ quantum dot embedded in monolithic microlenses.}
\end{figure}

In order to realize a fiber-coupled, ultra-compact, efficient and on-demand entangled photon pair source based on semiconductor quantum dots (QDs) it is necessary to efficiently couple the emission of the QDs operated at cryogenic temperatures to single mode fibers. This entails extracting the emitted light from the high refractive index matrix material semiconductors, such as $\GaAs$ exhibiting refractive indexes of about \num{3.5} \cite{Aspnes1986}, requiring engineering of the QD photonic environment due to the small angle of total internal reflection \cite{Shields2007}. Strategies for efficient extraction from semiconductors have been realized using various methods, such as open fiber, circular Bragg and photonic crystal cavities \cite{Tomm2021, Wang2019, Uppu2020}, as well as monolithic micro- and semiconductor solid immersion lenses \cite{Chen2018, Nie2021}. Secondly, the extracted light from the semiconductor quantum light sources needs to be efficiently inserted into a single mode fiber inside the cryogenic environment. This is necessary in order to keep the footprint of the system small and achieve compatibility to industrial environments - such as server rooms - because in these environments the employment of free-space optics needs to be avoided as vibration and temperature stability are limited.\\

It has been shown many times that $\GaAs$ QDs in $\AlGaAs$ matrix material are excellent sources of on-demand and maximally entangled photon pairs using their inherent exciton(X)-biexciton(XX) cascade using two-photon excitation scheme \cite{Stufler2006, Bounouar2015,Chen2018,Hopfmann2021}, see also \cref{fig:TPE}. In order to attain an efficient entangled photon pair sources from these QDs both X and XX emission needs to be coupled efficiently. Since the typical energy splitting of these emission lines is about \qty{3.6}{meV} \cite{Hopfmann2021a}, microcavities with high Q-factors, i.e. small bandwidths, cannot be used for efficient light extraction entangled photon pairs. For this reason, open-fiber cavities are not suitable for QD-based entangled photon sources, even though these systems have demonstrated very high extraction efficiencies of single photons from QDs into single mode fibers of up to \qty{57}{\percent} \cite{Tomm2021}. In previous work, we have shown that QD-nanomembranes attached to $GaP$ solid immersion lenses can be used to realize efficient sources of entangled photon pairs \cite{Chen2018}. However, due to the macroscopic nature they are fundamentally unsuited for direct fiber coupling inside a compact cryostat, as a macroscopic objective with high numerical aperture (NA) would be required to insert the entangled photon pairs into the single mode fiber. The latter can only be achieved in large cryostats unsuitable for operation in industry standard rack-systems. This leaves microscopic devices such as monolithic microlenses and circular Bragg cavities to achieve efficient extraction of entangled photon pairs. In this work, we chose the approach of monolithic microlenses as these are inherently broadband and polarization insensitive microscale devices, to achieve the same with circular Bragg cavities of moderate Q-factor would require very precise positioning of the QD with respect to the microcavity \cite{Wang2019, Liu2019}.\\

To provide efficient light extraction of the QDs from the $\AlGaAs$ matrix material, monolithic microlenses are used. A comprehensive and detailed analysis on the fabrication process, its optimization and performance statistics is found in our recent publication Ref. \cite{Langer2025a}. Further details can also be found in \cref{sec:sample_layout}.  The shape of the monolithic microlenses and the sample layout used in this work is shown explicitly in \cref{fig:Lens_shape} and \cref{fig:sample_layout}, respectively. It is worth noting that in the present work, the alignment between $\AlGaAs$ microlenses and QDs is not controlled. In order to achieve high extraction efficiencies, one has to postselect lenses with centrally embedded QDs. Based on the QD density of about \qty{0.2}{\per \um^2} and simulations, it can be estimated that the chance of finding a lens with a QD above \qty{80}{\percent} of the theoretical maximum extraction efficiency is about \qty{2}{\percent}. The identification of efficient QD entangled photon pair sources therefore requires characterization of many lenses using, for example, the hyperspectral imaging technique described in \cref{sec:Results}.\\

In order to tackle the challenge of efficient insertion of light emitted by single QDs into a single-mode fiber in a compact format at low temperatures, it is necessary to provide a high-NA optical system with small footprint. It has been previously demonstrated that 3D-printed micro-objectives can be fabricated by two-photon polymerization technique directly on the end of standard single mode fiber \cite{Gissibl2016}. Using this technique, it was possible to achieve rather complex designs with lensed fibers and 3D-printed lenses on QDs. This approach has been used to couple the emission of single $InGaAs$ QDs using 3D-printed objectives featuring NAs of circa \num{0.13} by gluing the components together \cite{Schlehahn2018, Bremer2020, Sartison2021}. While these achievements are considerable, the published single photon rate using pulsed excitation was limited to values below \qty{100}{\kilo \hertz}. The reason for this is that once glued, there is no possibility to adjust the mechanical alignment in any way \cite{Weber2024}, any temperature induced shift can therefore not be compensated for \cite{Schwab2022}. Furthermore, in order to collect more light emitted by a QD, coupling optics with higher NA are needed. So far, achieving high-NA 3D-printed micro-objectives requires complex designs with several lenses  \cite{Asadollahbaik2020}. These devices exhibit a significant drawback for coupling applications at cryogenic temperatures: such bulk 3D-printed structures are fragile and susceptible to temperature changes, especially because elements with different thermal expansion coefficients are combined. Recently, it has been shown that using no-core fibers is a promising approach to achieve high-NA 3D-printed micro-objectives on fibers using a single 3D-printed lens \cite{Ruchka2022}. Therefore, the design of the 3D-printed micro-objective optimized for \qty{780}{\nm} employed in this work uses the latter approach of only one aspheric lens in combination with a spliced no-core fiber of about \qty{500}{\um} length, see \cref{fig:Setup_SNOM}. By replacing the support structure (i.e., the 3D-printed expansion cylinder) with the piece of no-core glass fiber, the micro-objective becomes significantly more robust against mechanical shock and temperature drifts. Furthermore, we iteratively optimize the shape of the aspheric lens to compensate for the shape deviations, caused by two-photon polymerization 3D-printing. The combination of beam expansion in the no-core fiber piece with the highly aspheric optimized lens allows us to achieve NAs of up to \num{0.6}. Finally, instead of gluing the components together, fiber and sample mounts are engineered with the flexibility to be aligned using mechanical piezo stepper (range of \qty{4}{\milli \metre}) and scanner (range of \qty{30}{\micro \metre}) stages inside the cryostat. This provides the required flexibility to compensate for temperature drifts and enables long-term system stability. The overall experimental system design is further detailed in \cref{sec:Methods}.\\

\cref{fig:Setup_SNOM}a depicts the lensed fiber of type 780HP \footnote{Fiber diameter \qty{250}{\um} with polyamide coating of and \qty{125}{\um} bare.} is glued into a \qty{300}{\um} hole drilled into a standard \num{0.69}\,NA aspheric lens. The fiber with 3D-printed micro-objective is glued into the aspheric lens so that the tip of the fiber-objective has a distance of ca.\,\qty{1.5}{\milli \metre}, while the working distance of the lens is \qty{1.7}{\milli \metre}. The purpose of this arrangement is to be able to image the surface of the sample to find specific microstructures. Once found, the sample can be moved up by circa \qty{200}{\um} in order to collect the signal of the selected microstructure directly by the fiber micro-objective. When operated as an entangled photon pair source, the coarse imaging arrangement is inactive, it therefore does not affect the performance of the entangled photon pair source operation. In order to avoid any mechanical changes of the lensed fiber, it is fixed into the cryostat using strain relief and UV-glue.\\

\section{\label{sec:Results} Results and Discussion}

\begin{figure}
\includegraphics[width=0.5\textwidth]{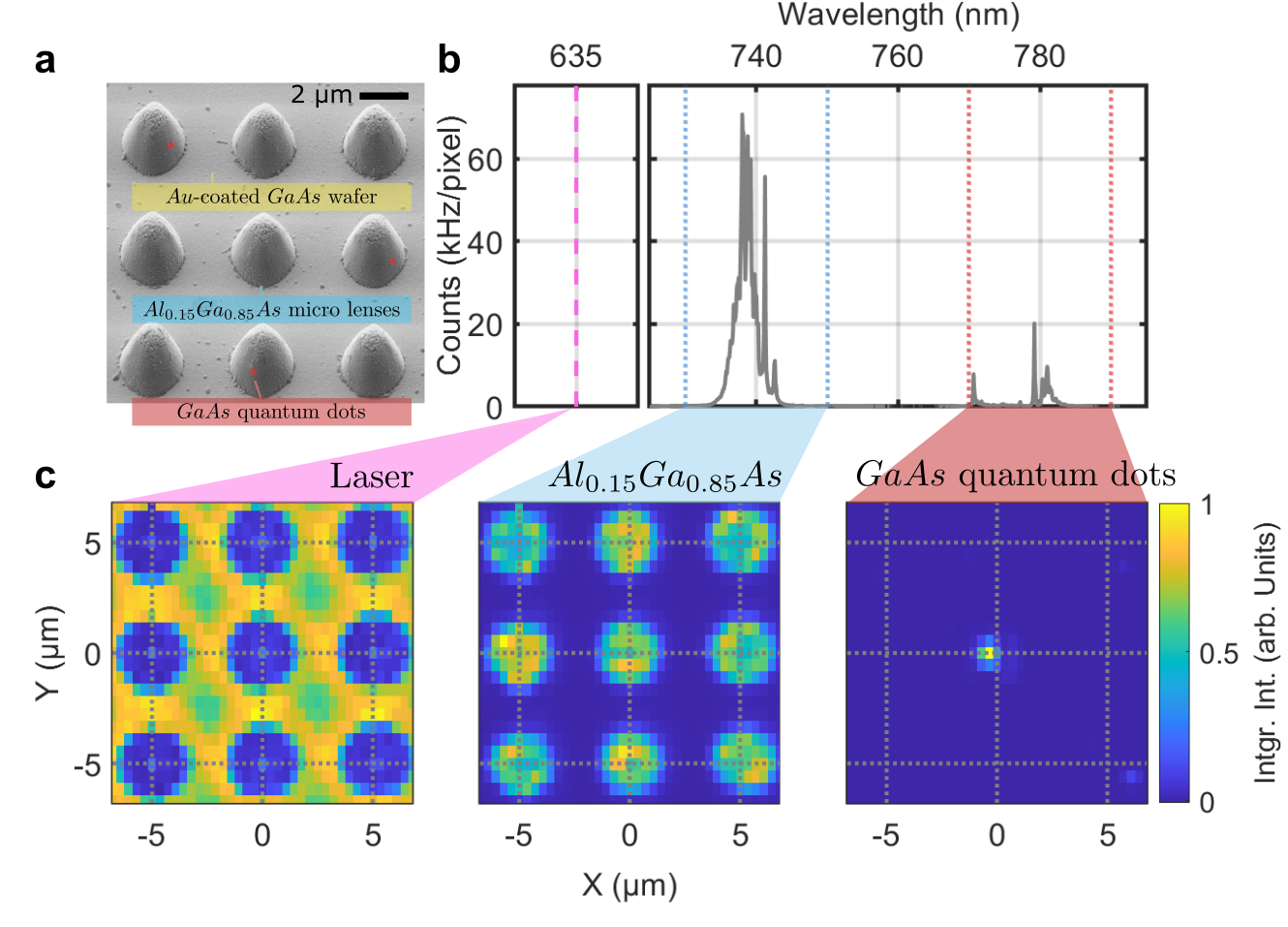}%
\caption{\label{fig:Lens_array} a) Scanning electron beam microscopy image recorded at an angle of \ang{45} of an array of $\AlGaAs$ monolithic microlenses on a gold-coated $\GaAs$ wafer. b) Exemplary photoluminescence spectrum using  \qty{635}{\nm} laser excitation collected through the single mode fiber using the 3D-printed micro-objective, cf. \cref{fig:Setup_SNOM}b. The wavelength bands at \qty{635 }{\nm}, \qtyrange{730}{750}{\nm}, and \qtyrange{770}{790}{\nm} of the excitation laser and luminescence of the $\AlGaAs$ matrix material and $\GaAs$ quantum dots are indicated in purple, light blue, and light red, respectively. c) Representation of the three distinct hyperspectral micrographs obtained by recording the different photoluminescence and reflection bands at wavelengths ($\lambda$) of the reflected laser, the $\AlGaAs$ monolithic microlenses and $\GaAs$ quantum dots, respectively, as a function of the spatial position using the XY-piezo scanner, cf. \cref{fig:Setup_SNOM}b. All micrographs are obtained simultaneously using the experimental apparatus detailed in Suppl. \cref{sec:Methods}.}
\end{figure}

To evaluate the usability and performance of the experimental design, and the scanning microlensed fiber in particular, the system is tested using hyperspectral imaging of a monolithic $\AlGaAs$ microlens array. By comparison of the nominal sample layout, cf. \cref{fig:sample_layout}, with high resolution scanning electron (SEM) beam microscopy images, cf. \cref{fig:Lens_array}a, with the wavelength- and position-resolved signal of the sample shown in \cref{fig:Lens_array}b and c, respectively, the spatial dimensions driven by the open-loop piezo scanners can be calibrated accurately. The experimental setup used for recording this hyperspectral imaging signal is detailed in \cref{sec:SNOM_Spec}. The laser reflection signal is maximal in-between the monolithic $\AlGaAs$ microlenses, while the luminescence signal of the microlenses (i.e., the band between \qtyrange{730}{750}{\nm}) behaves complementarily. For the QD emission band of \qtyrange{770}{790}{\nm}, only a fraction of the monolithic microlenses features centrally embedded QDs. On close inspection, one can find that some lenses feature weaker signal from non-centrally embedded QDs.\\

\begin{figure}
\includegraphics[width=0.5\textwidth]{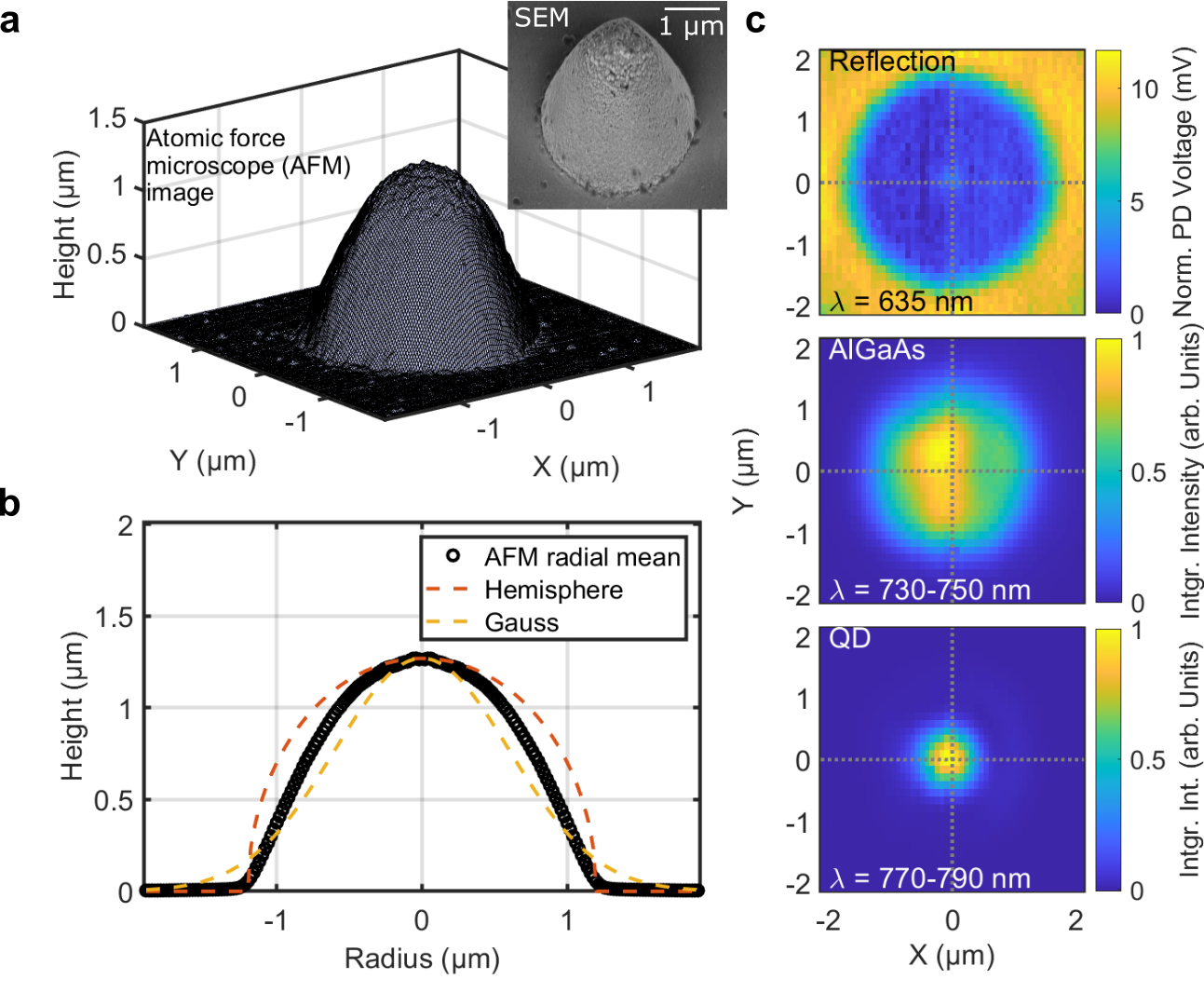}%
\caption{\label{fig:Lens_shape} a) Exemplary height profile of a single monolithic microlens with embedded quantum dots obtained using an atomic force microscope. Inset: Scanning electron beam micrograph recorded at an angle of a single microlens under an angle of \ang{45}. b) Height profile of the microlens of a) averaged along the lens azimuth, plotted as a function of its radius. The profiles of comparable hemispheric and Gaussian lens profiles are indicated by dashed lines. c) High resolution of the hyperspectral imaging micrographs for the different reflection and luminescence bands, cf. \cref{fig:Lens_array}. The full width at half maximum (FWHM) spatial resolution of the hyperspectral images collected through the single mode fiber is \qty{604 \pm 16}{\nm}, cf. Suppl. \cref{fig:resolution}.}
\end{figure}

In order to evaluate the limits of the optical and position-resolved performance of the scanning 3D-printed micro-objective at low temperatures, a detailed scan of a single microlens with embedded quantum dot is performed. This hyperspectral image together with comparable micrographs from atomic force (AFM) and SEM microscopy are shown in \cref{fig:Lens_shape}. Raw AFM and SEM  micrographs are shown in \cref{fig:Lens_shape}a, while the extracted radial profile of a microlens is depicted in \cref{fig:Lens_shape}b. One can deduce that the lens shape is in-between that of an ideal hemispherical and a Gaussian shaped lens. As discussed in great detail in Ref. \cite{Langer2025a}, the performance of the microlenses is best when they are close to a hemispherical shape, because this allows for the smallest possible angle of the emitted light of a centrally embedded QD at the semiconductor-air interface. Beam shaping of the emitted light into reduced emission angles (i.e., with $<1$ NA) should be regarded as a secondary goal of the monolithic microlens design. While the lens shape attained in the presented work is reasonably close to the hemispherical design, further improvements to the fabrication process would enhance the observed brightness further. The hyperspectral images shown in \cref{fig:Lens_shape}c corroborate the observations with the SEM and AFM microscopes very well. The determined lens diameter of \qty{3.0 \pm 0.3}{\um} in the reflection and $\AlGaAs$ bands demonstrate this nicely. If a QD is embedded centrally, as in this case, its spectral emission obtained in above-band excitation is enhanced by a factor of about \num{100} as compared to a quantum dot in an unprocessed $\AlGaAs$ matrix material. Note that both brightness and position of the QDs inside the lens can be imaged and characterized using the scanning microlensed fiber setup. The spectral resolution of the 3D-printed microlensed fiber at cryogenic temperatures is investigated using a QD in an unprocessed part of the QD-nanomembrane of the same sample, cf. \cref{fig:sample_layout}. The $\GaAs$ QDs in $\AlGaAs$ typically feature sizes of \qty{40}{\nm} and are therefore significantly smaller than the wavelength of light inside the material of ca. \qty{240}{\nm}. Therefore, the width of the resulting position dependent QD luminescence curve is in good approximation that of the point spread function of the 3D-printed micro-objective. Using this method, the FWHM resolution limit of the micro-objective is determined to \qty{604 \pm 16}{\nm}, cf. \cref{fig:resolution}. This value is within the expected theoretical diffraction limit of $d_\text{FWHM} = 0.51 \, \lambda/\text{NA} =$ \qty{663}{\nm} assuming the nominal numerical aperture of  \num{0.6} and wavelength of \qty{780}{\nm}. Note that the actual NA value can vary slightly for the fabricated micro-objective employed in this work, thereby explaining the mismatch of the observed to the theoretical limit. Nevertheless, it is clear that the 3D-printed micro-objective is operating close or even at the far field diffraction limit, thereby demonstrating the validity of the chosen design approach.\\

\begin{figure}
\includegraphics[width=0.5\textwidth]{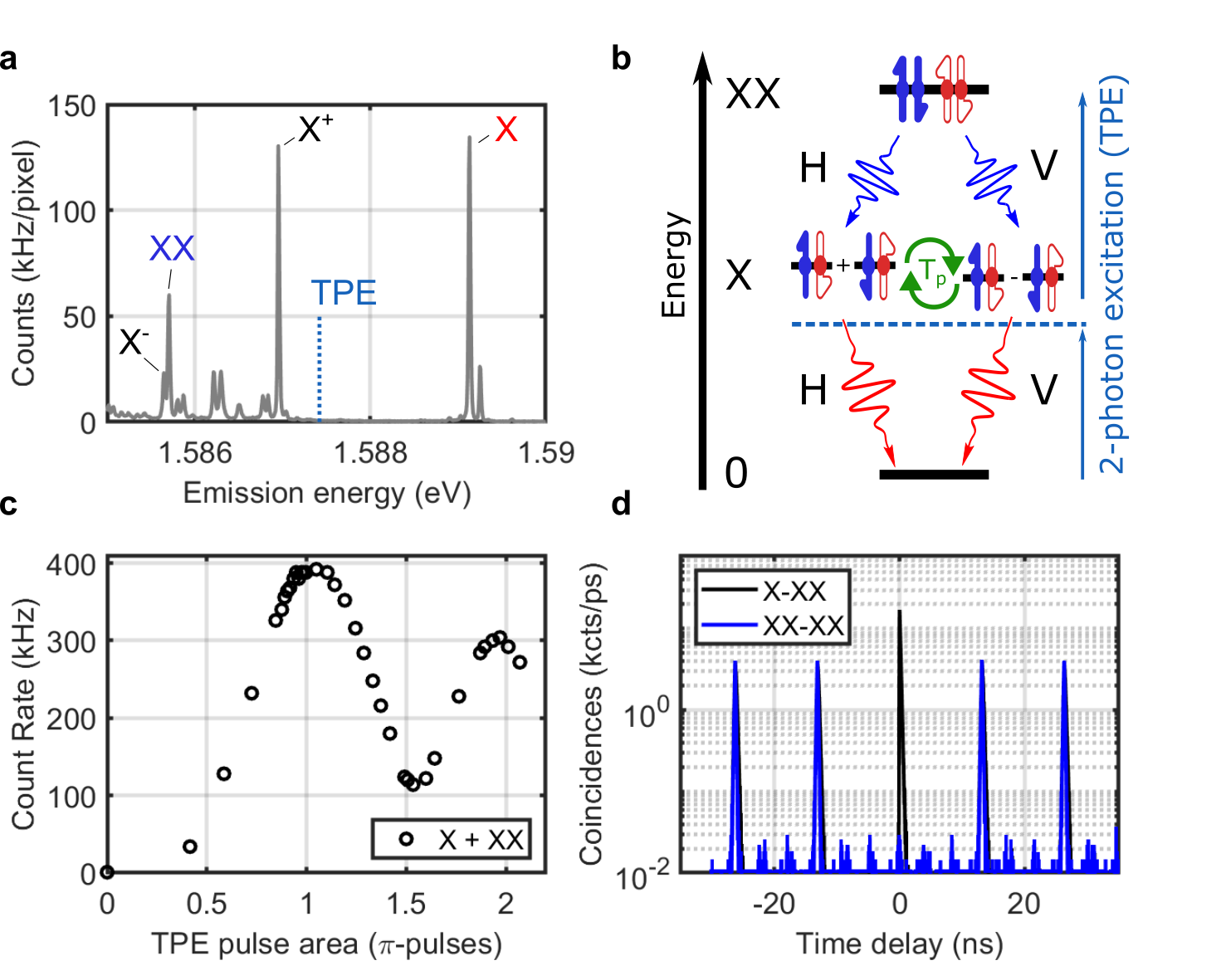}%
\caption{\label{fig:TPE} a) Above-band excitation spectrum of a single $\GaAs$ quantum dot as recorded through the 3D-printed single mode fiber objective. The different emission lines of the excitonic complexes of the exciton ($\X$), biexciton ($\XX$), and trion ($\Xm$) and ($\Xp$) as well as the energy of the resonant two-photon excitation (TPE) are annotated, cf. Ref. \cite{Hopfmann2021a}. b) Schematic illustration of the energetic level diagram of $\X$ and $\XX$ excitonic complexes used to obtain entangled photon pairs by the two-photon pulsed resonant excitation scheme. The $\X$ emission process precession is modulated by the precession period $T_p$, cf. Ref. \cite{Winik2017, Hopfmann2021}. c) Combined detector count rate of all four single photon detectors, cf. Suppl. \cref{sec:Exp_Setup}, as a function of the TPE pulse area. The first Rabi-$\pi$-pulse is achieved at a power of \qty{0.65}{\micro\watt} at a laser repetition rate of \qty{76}{\mega \hertz}. d) Measured coincidences versus photon arrival time delay in cross- and auto-correlation configuration of $\X$ and $\XX$, respectively, for polarization basis combination HH.}
\end{figure}

In order to determine the suitability of the combination of monolithic microlens with 3D-printed fiber collection optics as an ultra-compact entangled photon pair source, the bright QD, displayed in \cref{fig:Lens_shape}, is subjected to pulsed resonant two-photon excitation (TPE). This excitation scheme enables resonant driving of the $\XX$  via absorption of two photons with an energy in-between the $\X$ and $\XX$  transitions, see \cref{fig:TPE}a and b. Due to the coherent nature of this excitation scheme, Rabi-oscillations are observed as a function of the pulse power. This can be seen clearly in \cref{fig:TPE}c, where the principal Rabi-$\pi$-pulse energy is found to be \qty{8.5e-15}{\joule} at a \qty{76}{\mega \hertz} TPE rate. The observed combined X and XX single photon rate using superconducting single photon detectors is \qty{392 \pm 20}{\kilo \hertz}. This equates to a raw overall system per-pulse efficiency of \num{5.2 (3)e-3}. The single photon characteristics of in TPE is investigated using polarization resolved X and XX auto- and cross-correlation experiments. This is shown exemplarily for the H-H\footnote{For definition of the 6 polarization bases H,V,D,A,R and L see \cref{sec:prec_osc}.} polarization basis combination of emitted X and XX photons, respectively, in \cref{fig:TPE}d. In this graph, a clear bunching effect ($g_\text{X-XX}^{(2)}(0) \simeq 2.3$) for the X-XX cross-correlation and a strong anti-bunching in the XX-XX auto-correlation ($g_\text{XX-XX}^{(2)}(0) = $ \num{8.0 (6)e-4}) for time delays $\tau \to 0$ are observed.\\

\begin{figure}
\includegraphics[width=0.5\textwidth]{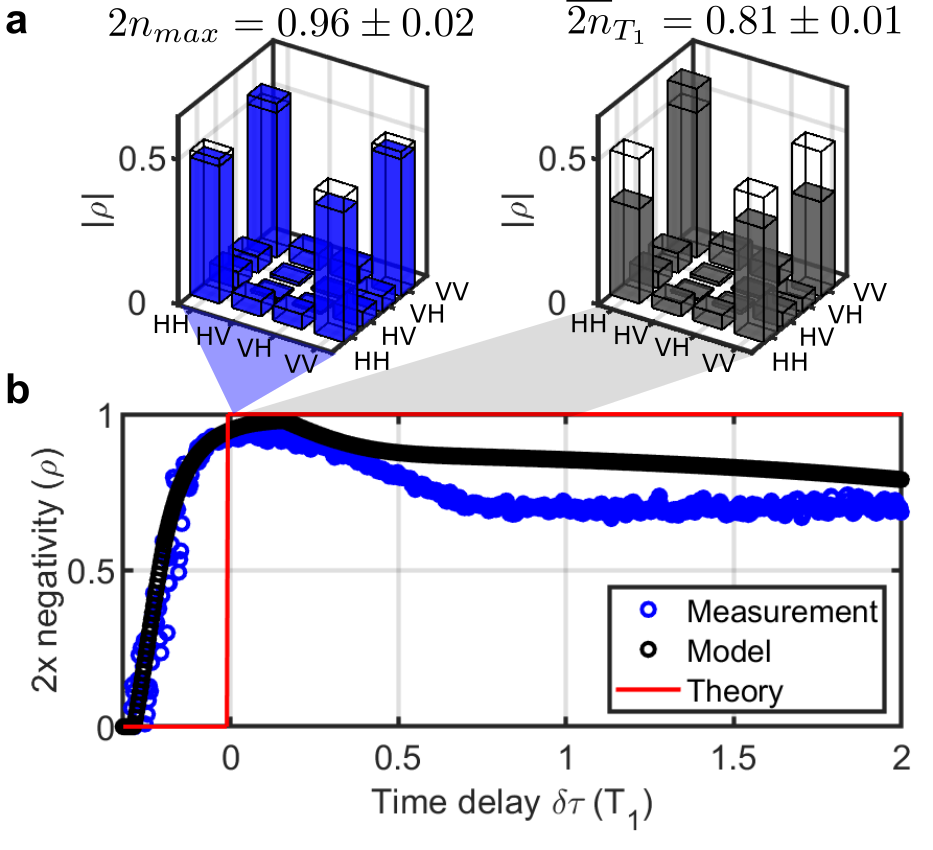}%
\caption{\label{fig:Ent_neg} a) Representation of the absolute value of extracted two-photon X-XX density matrixes $\rho$ using maximum likelihood estimation from the two-photon polarization tomography measurement, see also suppl. \cref{fig:corr_tomo}. Left: X-XX density matrix for maximal negativity value at time delay $\delta_\tau$ of $0$ using a bin width of \qty{4}{\pico \second}. Right: Average X-XX density matrix for emitted photons within one exciton lifetime $T_1^X = $ \qty{320 (1)}{\pico \second}, cf. suppl. \cref{fig:lifetime}. The respective entanglement negativity values $2n$ of both matrices are denoted above the charts. b) Entanglement negativity $2n$ as a function of time delay $\delta_\tau$ as extracted from the two-photon tomography measurement (blue), maximally entangled state model with detector timing resolution (black), and maximally entangled state theory (red), cf. suppl. \cref{sec:prec_osc}.}
\end{figure}

In the cascaded decay of the excited XX state, two photons, one at the X and XX transition energies, are emitted. Due to the preservation of the angular momentum in the emission processes from the initial XX state to the ground state $j_{\ket{XX}} = j_{\ket{0}} = 0$, the two emitted photons $j_{\nu} = \pm 1$ are polarization entangled. Because of the precession oscillations the X state exhibits due to its finite fine-structure splitting induced by spin-orbit coupling to its integer spin $j_{\ket{X}} = \pm j_{hh}\mp j_{e} = \pm \frac{3}{2} \mp \frac{1}{2} = \pm 1$ ($hh$ stands for heavy hole and $e$ for electron), the entanglement basis precesses around the H-V polarization axis. A detailed description of the creation of polarization entangled photon pairs from $\GaAs$ QDs and the precession effect can be found in the suppl. information and Refs. \cite{Winik2017,Hopfmann2021}. The fine-structure splitting $\Delta_{FSS}$ of this particular QD is determined to \qty{5.79 \pm 0.20}{\micro \electronvolt}, inducing a $\X$ state precession period $T_p^X$ of \qty{714 \pm 24}{\pico \second}. By measuring the full polarization tomography in all $6 \times 6 = 36$ polarization basis combinations of the time-resolved X-XX coincidences, the two-photon density matrix of the entangled state can be reconstructed. We are following the standard procedure as outlined, for example, in Ref. \cite{James2001}. The full tomography measurement, including an overlay of the data with a maximally entangled state model, is shown explicitly in suppl. \cref{fig:corr_tomo}. Due to the excellent agreement between the maximally entangled state model and the presented data, we conclude that the entangled state prepared by the presented entangled photon pair source is in good approximation maximally entangled. The model parameters $T_p^X$ and $T_1^X =$ \qty{320 (1)}{\pico \second} are determined in separate measurements, see \cref{sec:lifetime,sec:fss}. By comparison between the theory and ideal state curve with considering the detector time resolution, it can be concluded that the observable entanglement is mainly limited by the detector time resolution relative to $T_p^X$ and the precision of the polarization projection units, cf. \cref{fig:Ent_setup}, but not by the quality of the entangled source itself.  The extracted two-photon density matrices $\rho$ and the source entanglement negatives $2n$ are summarized in \cref{fig:Ent_neg}. The observed maximal entanglement negativity $2n$ over a \qty{4}{\pico \second} window is \num{0.96 \pm 0.02}, while the average over the photon pairs collected within one exciton lifetime $T_1^X$ equates to \num{0.81 \pm 0.01}. The extracted negatives as a function of time delay $\delta_\tau$ are depicted in \cref{fig:Ent_neg}b. The theoretical curve represents a maximally entangled state, see discussion in \cref{sec:prec_osc}, while the model is the same as the theory curve but convoluted with the detector time resolution. It is worth noting that the maximally entangled model is not fitted to the data, as it has no free parameters except the scaling to the coincidences. It is apparent that the entanglement of the emitted photon pairs levels off after an initial decrease from its maximum at about $2n = $ \num{0.7}, which remains intact for time delays of multiple $T_1^X$ and is only limited by the signal-to-noise ratio at $\delta \tau / T_1^X \gg 1$. While the maximally entangled model generally describes the observations quite well, the entanglement decay, mainly caused by time uncertainty in the measurement, exhibits some deviation. We attribute this to the uncertainties in the two-photon detection timing resolution $\delta_\text{FWHM}^\text{det} = $ \qty{126 (15)}{\pico \second} and exciton precession period $T_p^X$. Potentially, also effects which are not considered in the model and theory, such as spin relaxation during the $\XXtoX$-cascade, could play a contributing role in explaining the observed deviation. Nevertheless, the observations clearly demonstrate that the on-demand entangled photon pair source based on $\GaAs$ QDs embedded in monolithic microlenses and coupled to a 3D-printed micro-objectives on single mode fibers exhibits strong characteristics of a maximally entangled source even in the presence of fine-structure splitting.\\

\section{\label{sec:Conclusions} Conclusions}

We have demonstrated and verified a sophisticated and ultra-compact design of fiber-coupled semiconductor QD based entangled photon pair sources suitable for industrial applications. To achieve efficient and compact coupling of the emission of single QDs to a single mode fiber inside the cryostat at low temperatures of about \qty{3.8}{\kelvin}, while maintaining sufficient degrees of freedom to compensate temperature drifts. It is shown that a micro-objective 3D-printed directly on-top of a standard single mode fiber in combination with monolithic $\AlGaAs$ microlenses with embedded $\GaAs$ QDs is a viable approach to attain the desired entangled photon pair sources. Since fiber micro-objective and QD microlenses can be positioned against each other using a combination of piezo steppers and scanners, the systems retains a great deal of flexibility and is able to compensate for changes as well as can be employed for high resolution imaging. The latter is enabled by the diffraction-limited spatial FWHM resolution performance of \qty{604 \pm 16} {\nm} of the employed fiber micro-objective. Furthermore, this experimental system enables diffraction-limited hyperspectral imaging by utilizing the simultaneous reflection and luminescence signal of planar samples at cryogenic temperatures. By using a secondary capability to image the sample surface through a traditional microscopy setup co-aligned to the lensed fiber, the capability to efficiently find specific microstructures on a $mm$-sized sample is preserved.
In combination with the fabricated microlenses with centrally embedded QDs, single photon rates of up to \qty{392 \pm 20}{\kilo \hertz} at \qty{76}{\mega \hertz} pulsed resonant excitation are enabled. This constitutes about an order of magnitude improvement compared to investigations of similar systems \cite{Schlehahn2018, Bremer2020}. Additionally, using resonant two-photon excitation, this system can be employed to create entangled photon pairs featuring state-of-the-art entanglement negativities $2n$ of up to \num{0.96 \pm 0.02} and \num{0.81 \pm 0.01} when time intervals of \qty{4}{\pico \second} and $\X$-lifetime ($T_1^\X = $ \qty{320 \pm 1}{\pico \second}), respectively, are considered. The investigated QD quantum light source exhibits an uncorrected single photon purity of \qty{99.2 \pm 0.5}{\percent}. The observed two-photon density matrices are in good agreement with the model of maximally-entangled states, once the experimental limitations of the single photon timing jitter and accuracy of the polarization projection are accounted for.
In a next step, the presented ultra-compact sources will be integrated into mobile, standard \num{19} $in$ rack-systems in order to pioneer quantum networks in industrial environments.

\begin{acknowledgments}

We thank the clean room team, especially Martin Bauer and Ronny Engelhard, of the Leibniz IFW Dresden for their efforts and expertise in clean room processing of samples. The support of the Attocube Systems AG in the modification of the cryogenic system is gracefully acknowledged. This work was funded by the German Federal Ministry of Education and Research (BMBF) projects QR.X (contracts no. 16KISQ013 and 16KISQ016), QD-CamNetz (contract no. 16KISQ078), 3DprintedOptics (contract no. 13N14097). and Integrated3Dprint (contract no. 13N16875). H.G., P.R. and S.J. also acknowledge support from Baden-Wuerttemberg-Stiftung (project Opterial), European Research Council (Advanced grant Complexplas and PoC grant 3DPrintedOptics), German Research Foundation (DFG, grant no. 431314977/GRK2642), HORIZON EUROPE European Innovation Council (IV-Lab project with grant no. 101115545), Carl-Zeiss foundation (projects EndoPrint3D and QPhoton), and University of Stuttgart (Terra Incognita progame).

\end{acknowledgments}

\newpage\phantom{blabla}
\newpage\phantom{blabla}

\setcounter{figure}{0}
\makeatletter 
\renewcommand{\thefigure}{S\@arabic\c@figure}
\makeatother

\section{Supplemental Information}
\label{sec:Suppl}

\subsection{Methods}
\label{sec:Methods}

\subsubsection{Monolithic microlens sample}
\label{sec:sample_layout}

\begin{figure}
\includegraphics[width=0.5\textwidth]{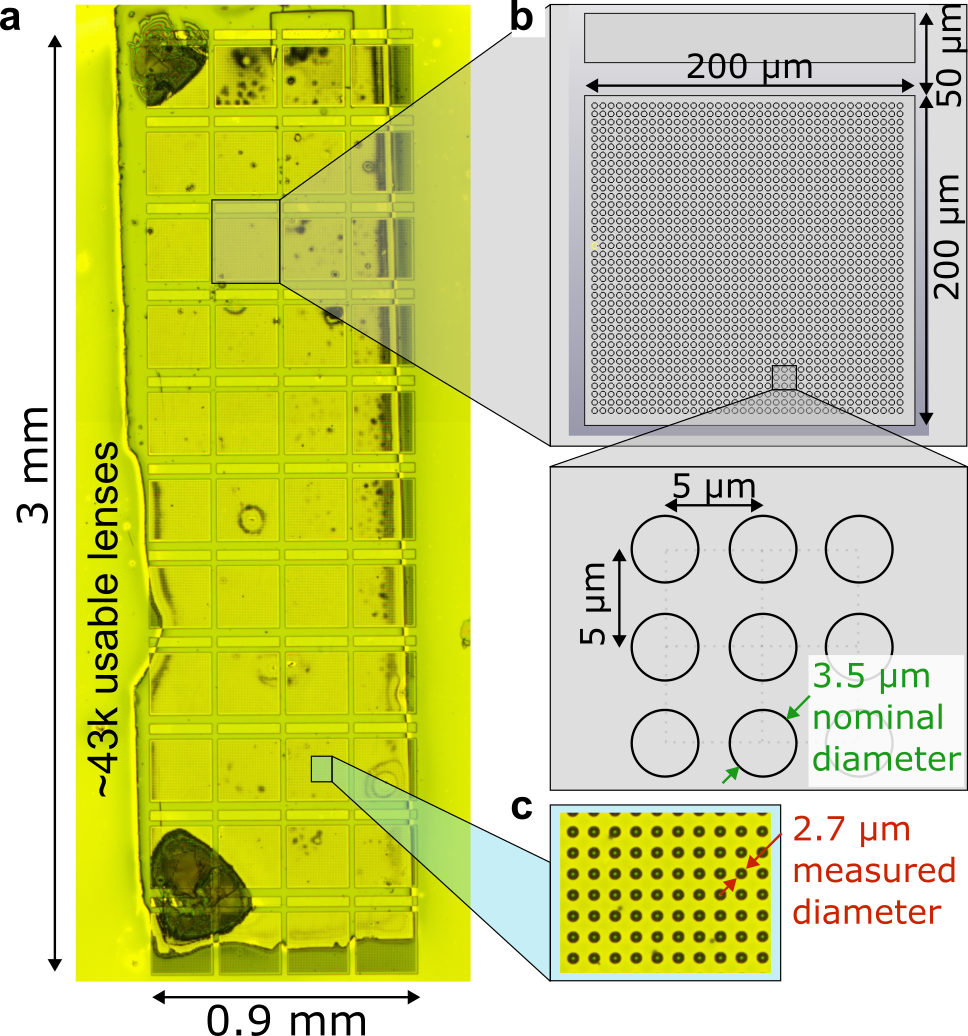}%
\caption{\label{fig:sample_layout} a) Microscopy image of the monolithic microlens sample used in this work. The membrane size on the gold-coated $GaAs$ waver is about \qtyproduct{3.0 x 0.9}{\milli\metre}, allowing the patterning of about \num{43000} lenses. b) Design of the optical lithography patterning used to fabricate the monolithic microlenses. Each patterning field contains $38 \times 38 = 1440$ mesas of \qty{3.5} {\micro \metre} nominal diameter. c) Close-up image of an array of fabricated monolithic microlenses using a temperature induced reflow and shape transfer process via reactive ion etching, see text and ref. \cite{Langer2025a} The shrinking of the diameter compared to the nominal design diameter is fully expected.}
\end{figure}

In the following, the fabrication process of the monolithic $\AlGaAs$ microlens sample used in this work is outlined. A comprehensive and systematic investigation of the manufacturing processes and the resulting optical performance of these microlenses is found in Ref. \cite{Langer2025a}.\\


Using molecular beam epitaxy, the $Al_{x}Ga_{1-x}As$ heterostructure is grown on a single crystal $3"$ $\GaAs$(100) wafer. After growth of a \qty{100}{nm} $\GaAs$ buffer an $\AlAs$-$\GaAs$ superlattice of \num{20} repetitions with layer thicknesses of \qty{3}{\nm} each, and a second buffer layer of \qty{100}{nm} $\GaAs$ are deposited. Consequently, an $\AlAs$ sacrificial layer of \qty{30}{\nm} width is grown. Next, an $Al_{0.15}Ga_{0.85}As$ layer of \qty{155}{nm} is deposited, followed by liquid droplet-etching of \qty{40}{\nm} wide and \qty{10}{\nm} deep nanoholes. Details of this process can be found in Ref. \cite{Keil2017}. The nanoholes are filled with \qty{2}{nm} of $\GaAs$ and overgrown by a capping layer of \qty{1500}{\nm} $Al_{0.15}Ga_{0.85}As$.\\

The obtained $Al_{x}Ga_{1-x}As$-heterostructure is processed into membranes with embedded QDs. This is performed in a photoresist-stabilized lift-off process by selective chemical etching of the $\AlAs$-sacrificial layer in a hydrofluoric acid solution. Consequently, large pieces up to several $mm^2$ are lifted off and transferred to \qty{70}{\nm} gold-coated $\GaAs$ single crystal wafer pieces. After dissolving the supportive photoresist and subsequent bake-out for enhanced bonding, a mostly flat $\AlGaAs$-membrane with embedded QDs is obtained. This is illustrated exemplarily in \cref{fig:sample_layout}a. The membrane has a clear gray contrast compared to that of the yellow-appearing gold substrate. On these membranes photoresist cylinders with nominal diameters of \qty{3.5}{\um} diameter and heights of \qty{0.38}{\um} are fabricated by standard photolithography techniques arranged in arrays of $38 \times 38$ cylinders, cf. \cref{fig:sample_layout}b. By heating the sample to \qty{140}{\degreeCelsius} for \qty{240}{\second}, the photoresist cylinders are reflowed into lens-shaped droplets. The latter is consequently transferred into the QD-membranes using chlor-argon-based reactive ion etching; see Ref. \cite{Langer2025a} for details of this process and the shapes obtained. Thereby the monolithic $\AlGaAs$ microlenses of about \qty{2.7}{\um} outer diameter and \qty{1.3}{\um} height, cf. \cref{fig:Lens_shape}, are realized. The shrinkage of the outer diameter of the etched $\AlGaAs$ compared to the photoresist lenses is fully expected. The sample features a total of \num{30} full fields of \num{1440} microlenses, yielding about \num{43000} lenses in total, of which about \qty{80}{\percent} are fabricated successfully, i.e. without significant defects and without inclusions between the membrane and the gold-coated $\GaAs$ substrate. Micrographs of the sample used in this work are depicted in \cref{fig:sample_layout}.\\

Since the location of QDs embedded in the membrane is random, only a small fraction of the microlenses feature bright quantum dots. By utilization of the QD surface density of typically \qty{0.1}{\um^2} and simulations of lateral misalignment of QDs and microlenses, see Ref. \cite{Langer2025a}, one can estimate that about 1 in every 200 lenses features QD emission intensities of \qty{>=50}{\percent} compared to a centrally embedded one. The yield of bright QD-microlenses could be increased to values of close to unity, if the photoresist cylinders, used for microlens fabrication, could be positioned accurately on top of single quantum dots. The latter could be achieved by integration of the described fabrication process with positioning of single lenses using methods such as cathodoluminescence lithography \cite{Gschrey2015}.

\subsubsection{Two-photon resonant excitation}
\label{sec:TPE}

It has been demonstrated that for the on-demand generation of entangled photon pairs from $\GaAs$ quantum dots, two-photon resonant excitation (TPE) is very suitable \cite{Stufler2006, Bounouar2015}. Utilizing this scheme, population inversion of the XX state can be achieved by $\pi$-Rabi-pulses. The Rabi oscillations as a function of resonant excitation pulse area are plotted in \cref{fig:TPE}b. For generation of the TPE pulses, a tunable optical parametric oscillator (OPO) pumped by a pulsed, frequency-doubled \qty{1032}{\nm} fiber laser with \qty{76}{\mega \hertz} repetition rate is used. The laser emission is filtered and shaped using a virtual Bragg notch filter and ruled reflection grating (\qty{1200}{lines\per \milli \metre}). The resulting spectral FWHM of the excitation pulses is about \qty{100}{\giga \hertz} around the QD TPE excitation resonance, which is located energetically between the $\X$ and $\XX$ emission energies, cf. \cref{fig:TPE}a. $\pi$-pulse excitation is achieved at an externally (i.e. before the fiber, cf. \cref{fig:Ent_setup}) measured power of \qty{0.65}{\micro \watt}, this corresponds to a $\pi$-pulse energy of \qty{8.5e-15}{\joule} per pulse.

\subsubsection{Polarization control}
\label{sec:PolControl}

Control over the detected polarization is achieved by employing two polarization projection (PP) units, consisting of a set of rotating $\lambda/2$ and $\lambda/4$ zero order wave plates optimized to \qty{780}{nm} before a polarizing beam splitter, see also \cref{fig:Ent_setup}. The PP units are calibrated and validated using a polarimeter and tunable continuous wave Ti:Sa laser tuned to the $\X$ and $\XX$ emission wavelengths for both detection paths of \cref{fig:Ent_setup}, respectively. Furthermore, the output polarizations of the PP units are rotated so that the QD polarization eigenbasis matches the laboratory eigenbasis. This enables the latter to be used to perform the two-photon correlation tomography (cf. \cref{fig:corr_tomo}) to extract the two-photon density matrix and investigate the entanglement of the emitted photon pairs, cf. \cref{fig:Ent_neg}. The average accuracy of the polarization projection units, when tested in the experimental scheme depicted in \cref{fig:Ent_setup}, is measured to be about \qty{96}{\percent} compared to the ideal projection. The reason for the deviation is that some of the optics between QD and PP units do not fully preserve the polarization due to the residual birefringence of the employed optics. These errors are the main reason why the measured entanglement, cf. \cref{fig:Ent_neg}, deviates slightly from that of a maximally entangled state.

\subsubsection{Two-photon correlation spectroscopy}
\label{sec:TPCS}

\begin{figure}
\includegraphics[width=0.5\textwidth]{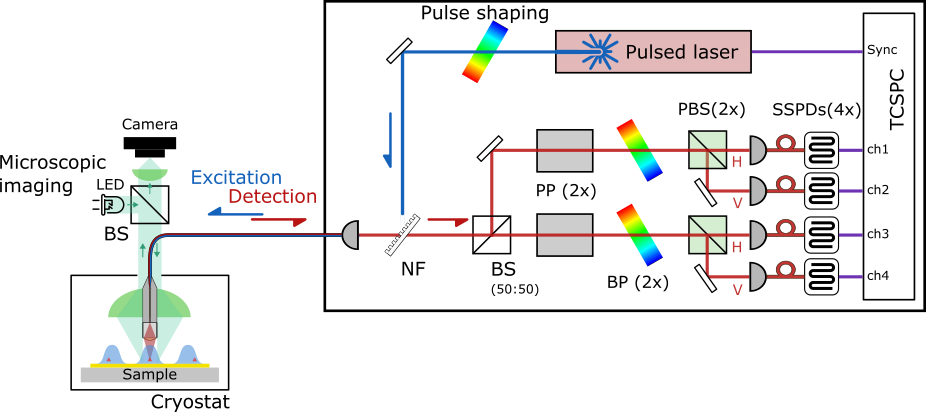}%
\caption{\label{fig:Ent_setup} Schematic illustration of the experimental apparatus used for characterizing the entanglement of the $\GaAs$ quantum dots excited by resonant two-photon excitation of the $\XX$, cf. \cref{fig:Lens_shape}. The polarization bases selection used in the two-photon correlation polarization tomography, cf. \cref{fig:corr_tomo}, is achieved by two polarization projection (PP) units consisting of a set of rotating $\lambda/2$ and $\lambda/4$ wave plates. The output polarization of the latter is projected onto a polarizing beam splitter (PBS), the signal of both output ports (i.e., the projections to H, V) of which are detected using superconducting single photon detectors (SSPDs). The electrical signal of these detectors is then correlated together with the laser sync signal in a time-correlated single photon counting (TCSPEC) device. The resonant two-photon excitation signal is separated from the detection signal using a volume Bragg grating notch filter (NF), the single photon signal is further filtered by transmission gratings acting as band pass (BP) filters. The free-space imaging microscopy CMOS-camera, LED-illumination, and beam splitter (BS) used for course imaging of the sample in the cryostat is mounted directly on top of the cryostat.}
\end{figure}

Two-photon correlation signal is obtained by correlating the arrival time between detection events of the four superconducting single photon detectors (SSPDs). The electronic correlation is achieved by an electronic correlation unit, which groups the resulting coincidences by the time difference between the detection channels. Since the projected polarization can be chosen independently only for two of the four channels (cf. \cref{fig:Ent_setup}), the dependent polarizations coincidences can be added to the respective polarization combinations. For example, when both PP units are set to the combination HH (in transmission of the respective polarizers) the two-photon events recorded from the remaining combinations of output ports will be HV, VH and VV. These coincidences can then be added to the respective polarization combinations of the tomography measurement, cf. \cref{sec:corr_tomo}. The FWHM timing jitter $\delta_{1P}$ of each detector is determined to about \qty{89}{\pico \second} by modeling the single photon detection events generated by the pulsed laser relative to the laser sync signal to the system to a normal distribution. It follows that the two-photon coincidence FWHM timing jitter $\delta_{2P} = \sqrt{2}\, \delta_{1P}$ is $ \simeq $ \qty{126}{\pico \second}. To fully suppress the excitation laser in the single photon signal, a single notch filter is not sufficient. Therefore, each detection path features an additional transmission grating (\qty{800}{lines \per \milli \metre}) acting as \qty{75}{\giga \hertz} FWHM band pass filters for both $\X$ and $\XX$ detection paths.

\subsubsection{High-resolution photoluminescence spectroscopy}
\label{sec:SNOM_Spec}

\begin{figure}
\includegraphics[width=0.5\textwidth]{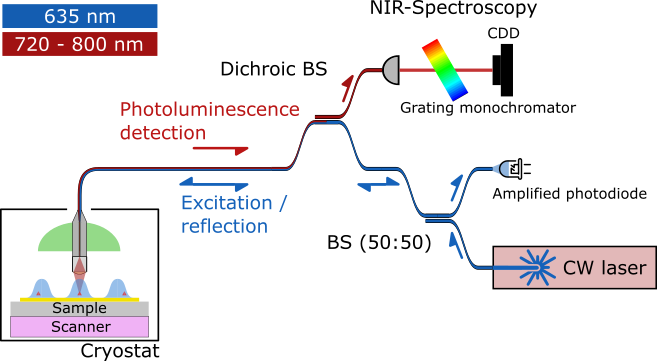}%
\caption{\label{fig:SNOM_Spec} Schematic illustration of the experimental apparatus used for the hyperspectral spatial imaging shown, cf. \cref{fig:Lens_array,fig:Lens_shape}. The positioning control in the horizontal xy-plane is provided by a calibrated open loop piezo scanner. Vertical Z-focusing and coarse xy-position control of the sample is achieved by a xyz-piezo stepper, cf. \cref{fig:Setup_SNOM}b. The laser and spectroscopy signal detected by a charge coupled device ($CCD$) after a grating monochromator is split using a fiber coupled dichromatic beam splitter ($BS$). The continuous wave ($CW$) excitation laser and its reflected signal detected by an amplified photodiode is separated by a fiber coupled $50:50$ beam splitter. We use single-mode 780HP fibers.}
\end{figure}

The spectroscopic signals presented in this work are obtained using a double-stage grating spectrometer of \qty{0.75}{\metre} focal length and an attached charge coupled device (CCD). The spectroscopy system can be used with different ruled gratings (\num{300}, \num{1200}, and \qty{1800}{lines \per \milli \metre}) in single and double stage configuration. In the configuration of the highest spectral resolution, a FWHM of \qty{15}{\micro \electronvolt} is achieved. The input signal of the spectroscopic system is fiber-coupled, see \cref{fig:SNOM_Spec}. This system is therefore not only used for spatially resolved hyperspectral imaging, using above-band excitation and piezo scanners, but also to verify the TPE pulse shape and spectral filtering of the detection channels using the notch filters and transmission gratings, cf. \cref{sec:Exp_Setup}.

\subsubsection{Modelling of maximally entangled photon pair coincidences}
\label{sec:prec_osc}

\begin{figure}
	\includegraphics[width=0.5\textwidth]{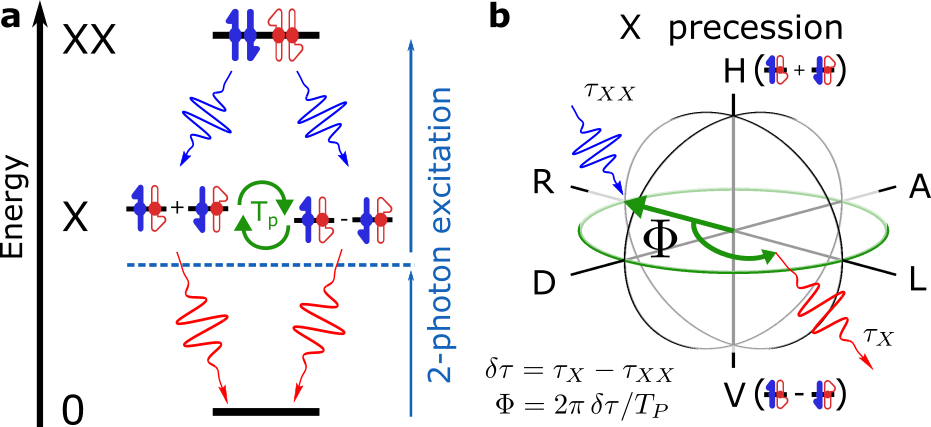}	
	\caption{\label{fig:X_precession} Schematic illustration of the exciton-biexciton ($\XXtoX$)-cascade. (a) Decay using an energetic level scheme which illustrates the polarization precession (b) with oscillation period $T_P$ of the $\X$ state due to its finite fine structure splitting $\FSS = h/T_P$. The precession only affects polarization states which are not in the H-V eigenbases. The phase $\Phi$ in the R-L and D-A polarization plane acquired by the second ($\X$) with respect to the first ($\XX$) photon of the cascade is therefore given by the difference of the $\XX$ and $\X$ decay times $\delta \tau$ in relation to $T_P$: $\Phi = 2 \pi \delta \tau / T_P$.}
	
\end{figure}

The modeling of maximally entangled photon pair sources with finite fine-structure splitting, which includes a formalism of the $\X$ precession oscillations, presented in this section, is directly adapted from \emph{Winik et. al.} \cite{Winik2017} as well as Ref. \cite{Hopfmann2021}.

Due to the finite energetic $\X$ finestructure splitting between its $\ket{H}$ and $\ket{V}$ eigenstates, the initial state of $\X$ after decay of $\XX$ will acquire a phase $\Phi$ over time. This effect leads to a precession oscillation observable in the D, A, R, and L polarization bases and is illustrated in \cref{fig:X_precession}. The two photon state originating from the $\XXtoX$-cascade can therefore be expressed in the $\X$ eigenbases by

\begin{align}
	\label{eq:X_precession_state}
	\ket{\Psi_{P_1,P_2}(\tau)} = \frac{1}{\sqrt{2}} (\ket{H_1 H_2} + e^{-i 2 \pi \delta \tau \FSS/h}\ket{V_1 V_2}) \,,
\end{align}

where the first photon corresponds to the one emitted from the biexciton, the second to the exciton emission and $\delta \tau = \tau_{\X} - \tau_{\XX}$ is the time difference between the two events. The precession oscillation period is given by $T_p = h/\FSS$ with $h$ as the Planck constant. It follows that the time dependent two-photon density matrix $\rho_{P_1,P_2}(t)$ in the $\ket{H}$ and $\ket{V}$ eigenbases can be expressed as

\begin{align}
	\label{eq:DM_precession_state}
	\rho_{P1,P_2} = \rho =\frac{1}{2} \left(
	\begin{array}{cccc}
		1 & 0 & 0 & e^{-i 2 \pi \delta \tau \frac{\FSS}{h}} \\
		0 & 0 & 0 & 0 \\
		0 & 0 & 0 & 0 \\
		e^{i 2 \pi \delta \tau \frac{\FSS}{h}} & 0 & 0 & 1 \\
	\end{array}
	\right) \, .
\end{align}

The entanglement negativity $\Neg$ of this density matrix is independent of $\tau$, i.e., the two-photon state is always maximally entangled $\Neg(\rho_{P1,P_2}) = \left|e^{i 2 \pi \delta \tau \frac{\FSS}{h}}\right|/2 = \frac{1}{2}$.
 
The six orthogonal polarization bases (H, V, D, A, R, L) are mutually interdependent and span the polarization Poincaré sphere, c.f. \cref{fig:X_precession}(b). The bases D, A, R and L can therefore be expressed as complex superpositions of $\ket{H}$ and $\ket{V}$: $\ket{D} = \frac{\ket{H} + \ket{V}}{\sqrt{2}}$, $\ket{A} = \frac{\ket{H} - \ket{V}}{\sqrt{2}}$, $\ket{R} = \frac{\ket{H} - i \ket{V}}{\sqrt{2}}$ and $\ket{L} = \frac{\ket{H} + i \ket{V}}{\sqrt{2}}$, respectively. In terms of polar coordinates $P(\theta, \phi) = \text{cos}(\theta/2) \ket{H} + e^{i \phi} \text{sin}(\theta/2) \ket{V}$ of the Poincaré sphere we obtain: $P_H = (0, \phi)$, $P_H = (\pi, \phi)$, $P_D = (\frac{\pi}{2}, 0)$, $P_A = (\frac{\pi}{2}, \pi)$, $P_R = (\frac{\pi}{2}, \frac{3 \, \pi}{2})$ and $P_L = (\frac{\pi}{2}, \frac{\pi}{2})$.

The polarization-resolved two-photon correlation traces $C_{IJ}^{\text{2P}}(\tau)$ with bases combinations $IJ$ are proportional to the conditional probability of detecting one photon from the $\XX$ and one from $\X$ transition at a later time $\tau$ and are given by

\begin{align}
	\label{eq:X_XX_two_photon_corr_theo}
	\begin{split}
	& C_{IJ}^{\text{2P(theory)}}(\delta \tau) \propto p_{\X}\left| \braket{P_I,P_J}{\Psi_{P_I,P_J}(\delta \tau)}\right|^2 \\	
		& = \frac{e^{- \frac{\delta \tau}{T_1^\X}}}{2 \, T_1^\X} \left| \text{cos} \frac{\theta_I - \theta_J}{2} \, \text{cos} \left( \frac{\phi_I + \phi_J}{2} + \frac{\pi \, \delta \tau \, \FSS}{h} \right) \right. \\
		& \left. + i \, \text{cos} \frac{\theta_I + \theta_J}{2} \, \text{sin} \left( \frac{\phi_I + \phi_J}{2} + \frac{\pi \, \delta \tau \, \FSS}{h} \right) \right|^2 \, .
	\end{split}
\end{align}

Note that only the 16 polarization combinations in non-HV-eigenbases, i.e., $\theta_I = \theta_J = \pi/2$,  exhibit precession oscillations. For all other combinations the oscillatory term reduces to a constant and is therefore proportional to a simple lifetime measurement of $\X$ - see also \cref{sec:lifetime}. Equation (\ref{eq:X_XX_two_photon_corr_theo}) corresponds to the theory curve of \cref{fig:corr_tomo}.\\

In order to model the recorded two-photon correlation traces, the limited time-resolution of the detection system needs to be taken into account. This is carried out by analytical convolution of $C_{IJ}^{\text{2P(theory)}}(\delta \tau)$ with the two-photon event timing jitter distribution $\DetResPP{2P}$:

\begin{equation}
	\label{eq:X_XX_two_photon_corr_model}
	C_{IJ}^{\text{2P(model)}}(\delta \tau) = C_{IJ}^{\text{2P(theory)}}(\delta \tau) \, \ast \, \DetResPP{2P} \, ,\\
\end{equation}

where $\DetResPP{2P}$ represents a normal distribution with a FWHM of the two-photon coincidence timing jitter $\delta_{2P}$. This function is plotted in \cref{fig:corr_tomo} as the model of the two-photon coincidence curve and has no free parameter (besides the y-axis scaling). It is worth pointing out that this model still assumes a maximally entangled two-photon state.

\subsubsection{3D-printed micro-objective fabrication}
\label{sec:obj_fabriaction}

To design the micro-objective we use commercially available software Ansys Zemax OpticStudio. In our work, the micro-objective comprises a singlet aspheric lens with a numerical aperture of 0.6. 

As described in the main text, we use the two-photon polymerization 3D-printed technique to fabricate the micro-objective. Before printing the micro-objective on a fiber, we perform a so-called iterative optimization using confocal microscope (µsurf expert, NanoFocus AG). For this, we print two of the designed micro-objective on a glass substrate and measure them with the confocal microscopy. After extracting vertical and horizontal topography profiles of these two lenses, we average them and calculate the deviations between the initial design and the measured shape. Afterwards, the derived deviations are added to the design, to form the first iterated design. Then, this design is printed in the same fashion as the initial one. We repeat the protocol of printing, measuring and compensating the lens profile inaccuracies until we obtain profile deviations under 1 µm. 
Afterwards, we can 3D-print the shape-optimized micro-objective directly onto the facet of the single-mode fiber of type 780HP.

 In order to do so, the fiber is manually stripped, cleaved and spliced with a so-called no-core fiber (FG125LA, Thorlabs GmbH) using automated glass processor Vytran GPX3800. In the same step, we cleave the spliced fibers to obtain a single piece of the 780HP with a \qty{500}{\nm} long no-core fiber on its end. After splicing, we mount the prepared fiber in the 3D printer Nanoscribe Photonic Professional GT (Nanoscribe GmbH, Karlsruhe) and align the system on the facet of the no-core fiber. The printing process for the optimized micro-objective is \qty{30}{\min}. When the printing is finished, we develop the fiber in mr-Dev600 solvent (micro-resist technology GmbH) for 15 min, and rinse with 2-isopropanol for 3 min. Finally, we expose the 3D-print to the UV-light for 5 min to ensure the full polymerization and the homogenity of the material inside the micro-objective.

\subsection{Spatial resolution of lensed single mode fiber}
\label{sec:resolution}

\begin{figure}
\includegraphics[width=0.5\textwidth]{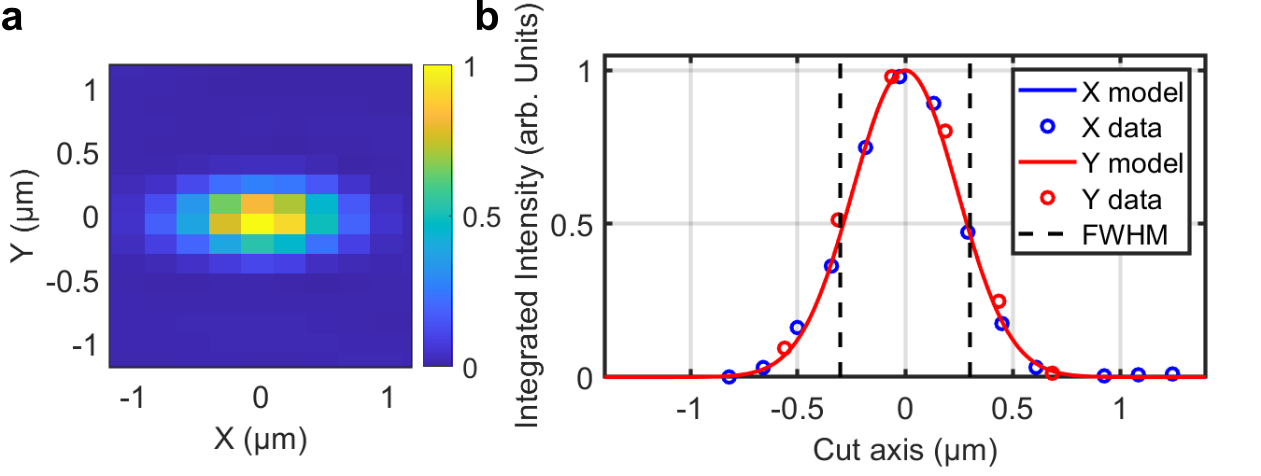}%
\caption{\label{fig:resolution} a) Integrated spectral image in the \qtyrange{770}{790}{\nm} wavelength band of a single quantum dot in an unprocessed membrane (i.e. no microlens). Note that the $x$- and $y$-pixel spacing is not equivalent. b) Cross-section of a). The full width at half maximum (FWHM) of \qty{604 \pm 16}{\nm}, obtained by modelling of the cross-sections with a normal distribution, is indicated by black dashed lines.}
\end{figure}

In order to evaluate the practically achievable spatial resolution of the scanning micro-objective on a single mode fiber at cryogenic temperatures, it is useful to experimentally determine the point spread function (PSF) of this system. To do this practically, a point-like defect on the actual sample can be imaged through the fiber. The QDs are a good approximation of such a point-like defect, as their typical lateral size of about \qty{40}{\nm} is significantly smaller than the wavelength in the $\AlGaAs$ matrix material of about \qty{230}{\nm} \cite{Keil2017}. In order to assure that the microlenses do not influence the measurement of the PSF, a QD on the unprocessed membrane between the microlens fields, see \cref{fig:sample_layout}, is imaged using the experimental setup described in \cref{sec:SNOM_Spec}. The resulting integrated position-resolved spectrum in the band of \qtyrange{770}{790}{\nm} is depicted in \cref{fig:resolution}a. The extracted cuts along the horizontal and vertical axis are presented in \cref{fig:resolution}b. Following the argument above, these cuts represent in good approximation the PSF in these directions and are modeled by a normal distribution exhibiting a FWHM of \qty{604 \pm 16}{\nm}. No discernible squeezing of the PSF in either the vertical and horizontal spatial directions is observed, indicating a radially symmetric operation of the 3D-printed micro-objective ontop of the single mode fiber.

\subsection{Lifetime}
\label{sec:lifetime}

\begin{figure}
\includegraphics[width=0.5\textwidth]{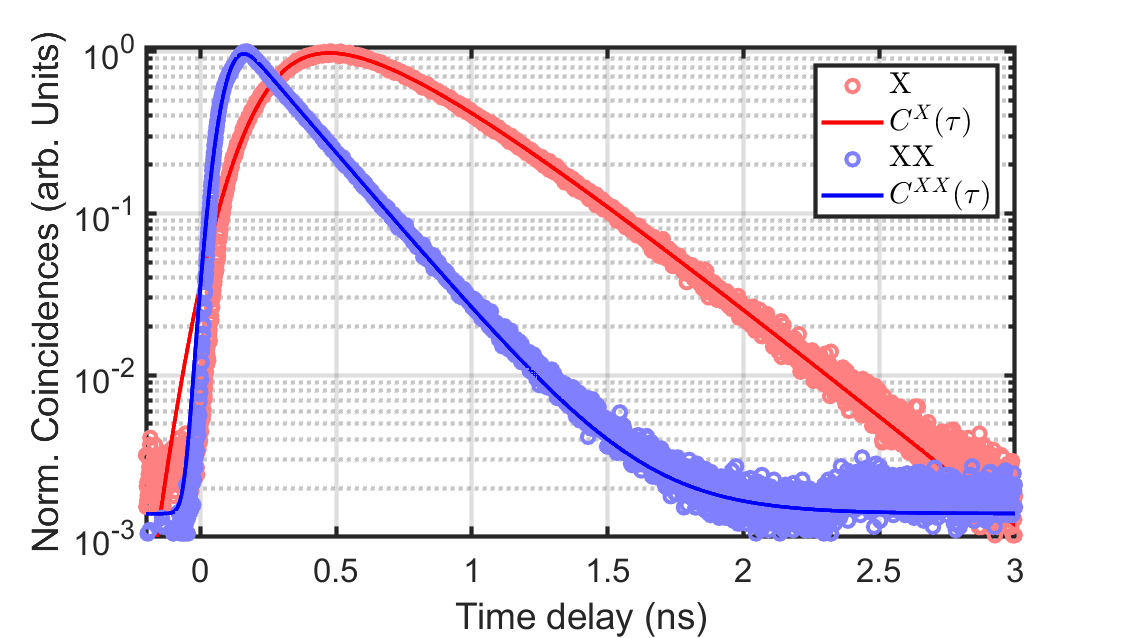}%
\caption{\label{fig:lifetime} Single photon coincidences as a function of time delay with respect to the sync signal of the two-photon resonant excitation laser pulses, cf. Fig. 3. The experimental data of the ($\X$) $\XX$ emission is modeled to (delayed) exponential decay curves $C^{\X}(\tau)$ and $C^{\XX}(\tau)$, respectively, see text for details.}
\end{figure}

The lifetime of QD transitions is typically inferred by recording the time-resolved photon emission correlation traces triggered by resonant excitation pulses. We employ two-photon correlation data recorded in the H polarization base using $76$ MHz TPE to determine the respective $\X$ and $\XX$ lifetimes. The resulting correlation traces are depicted in \cref{fig:lifetime}. The corresponding correlation functions models $C(\tau)$ are given by

\begin{align}
	C^{\XX}(\tau) &\propto e^{-\tau/T_1^{\XX}} \ast \DetResP{1P} \, , \\
	C^{\X}(\tau)  &\propto e^{-\tau/T_1^{\XX}} \ast e^{-\tau/T_1^{\X}} \ast \DetResP{1P} \, . 
\end{align}

The single photon detection event time delay is given by $\tau$. The only free modeling parameters are the lifetimes, which equate to $T_1^{\X} = $ \qty{320 \pm 1}, $T_1^{\XX} = $ \qty{222.7 \pm 0.7}{\pico \second}.\\

\subsection{Exciton fine structure splitting}
\label{sec:fss}

\begin{figure} %
\includegraphics[width=0.5\textwidth]{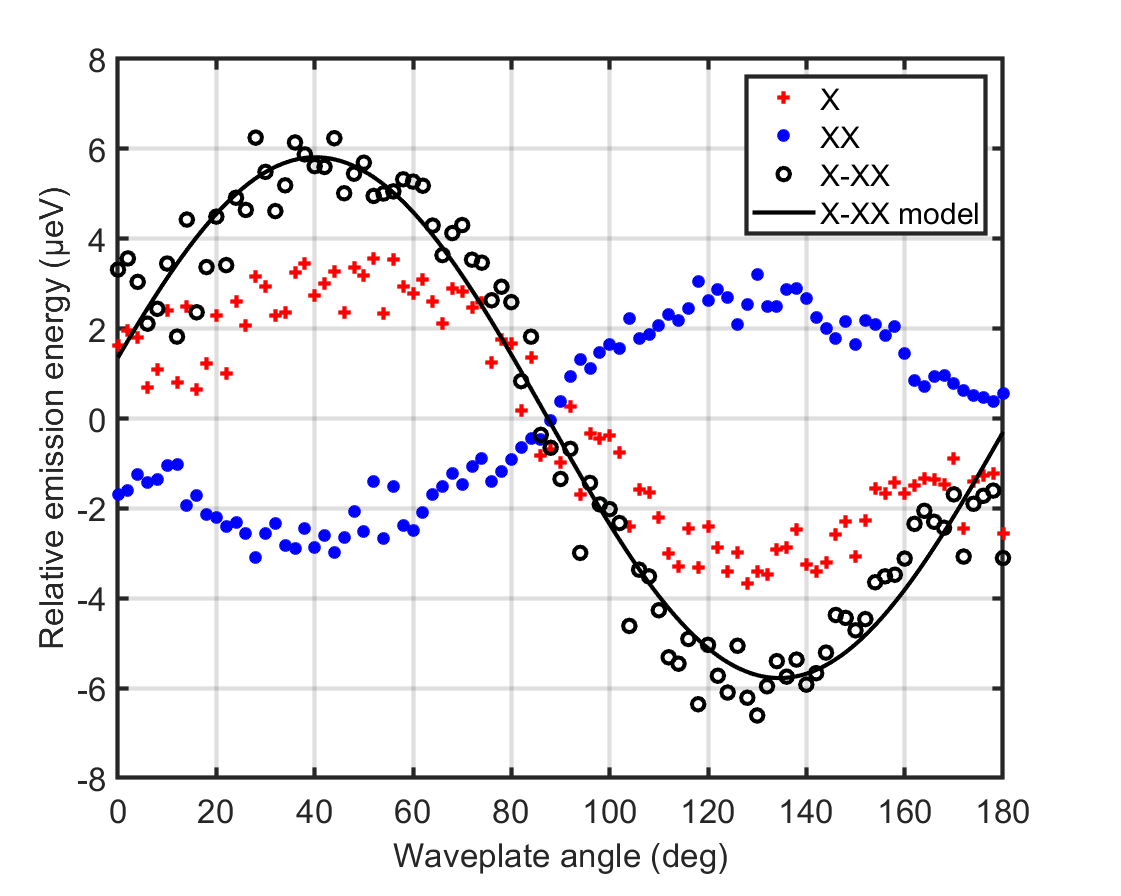}%
\caption{\label{fig:fss} Variation of $\X$ and $\XX$ emission line energies as a function of the $\lambda/2$ wave plate angle of the polarization projection unit, see \cref{fig:Ent_setup}. The energy variation is shown for both $\X$ (red), $\XX$ (blue) and their difference $\X$-$\XX$ (black). The latter is modeled, see text for details, yielding an excitonic fine-structure splitting $\Delta_{\FSS}$ of \qty{5.79 \pm 0.2}{\micro \electronvolt}.}
\end{figure}

The $\X$ fine structure splitting ($\FSS$) of $\ket{H} = \frac{1}{\sqrt{2}} (\ket{\uparrow \Downarrow} + \ket{\downarrow \Uparrow})$ and $\ket{V} = \frac{1}{\sqrt{2}} (\ket{\uparrow \Downarrow} - \ket{\downarrow \Uparrow})$ is revealed by its characteristic linear polarization dependence on the central emission energy \cite{Bayer2002}. This can experimentally be realized by recording PL emission spectra as a function of the angle of a lambda half plate followed by a fixed polarizer, cf. \cref{fig:Ent_setup}. The relative emission energies of exciton ($\X$), biexciton ($\XX$), and their difference ($X-XX$) of such an experiment are depicted in \cref{fig:fss} together with respective modeling to a sine function. From the latter the fine structure splitting $\FSS(\X)$ of \qty{5.79 \pm 0.2}{\micro \electronvolt} is obtained.\\

\subsection{Correlation tomography}
\label{sec:corr_tomo}

\onecolumngrid

\begin{figure}[htb]
\includegraphics[width=1.0\textwidth]{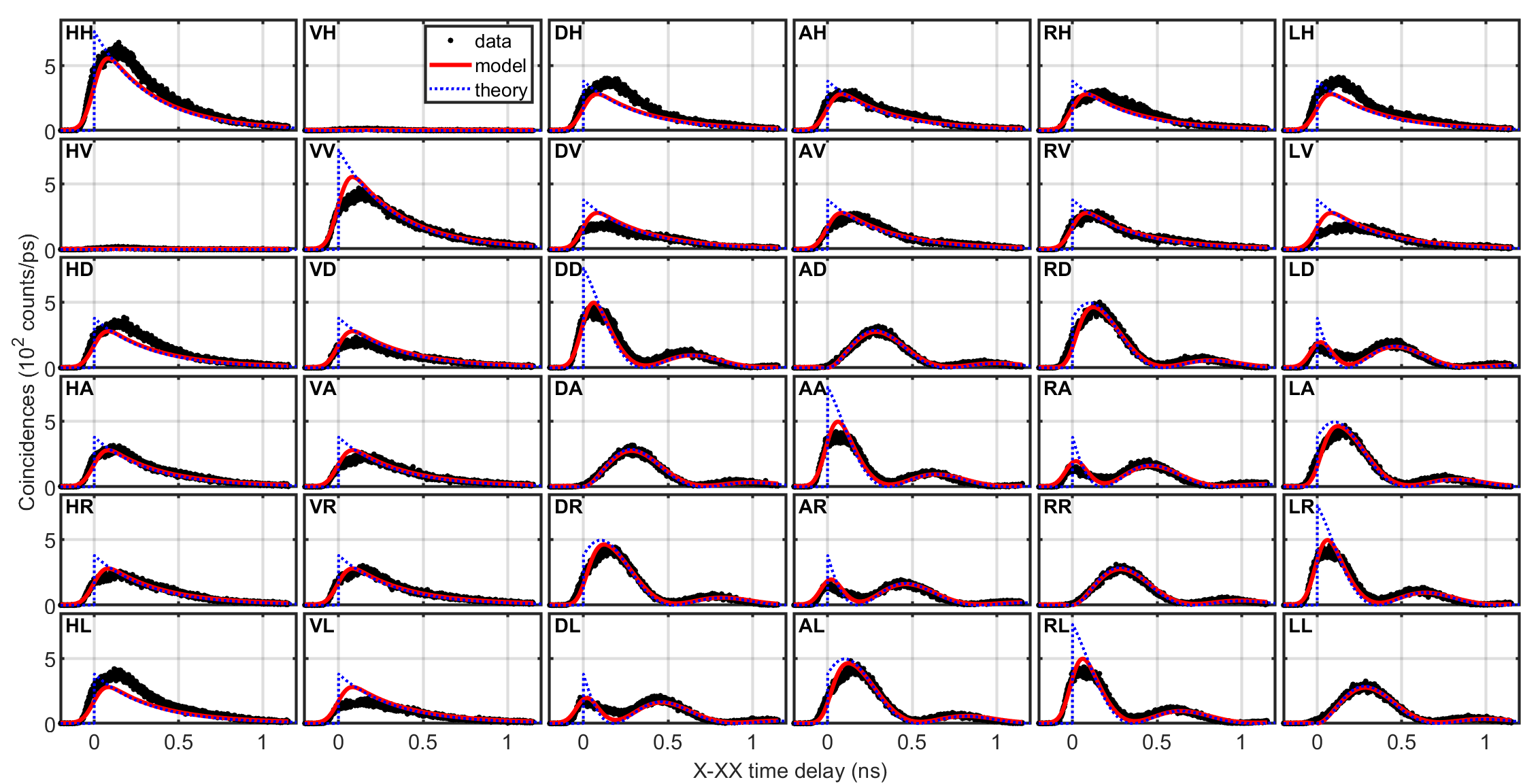}%
\caption{\label{fig:corr_tomo} Two-photon $\X$-$\XX$ coincidence polarization tomography as a function of the $\X$-$\XX$ emission time delay. The raw data is shown as black dots. The maximally entangled state theory (blue dotted line) and model (red line), see \cref{sec:prec_osc} and specifically \cref{eq:X_XX_two_photon_corr_theo,eq:X_XX_two_photon_corr_model} for details, does not have any free parameters, except the coincidence axis scaling. The remaining parameters, such as $T_1^\X$ and $\Delta_{FSS}$, are obtained by separate measurements, cf. \cref{sec:lifetime,sec:fss}.}
\end{figure}

\twocolumngrid

$\XXtoX$ two-photon correlation traces recorded by TSPC histograms of \qty{1}{ps} bin width are depicted in  \cref{fig:corr_tomo} for different polarization base combinations. The raw two-photon coincidence data is modeled to both maximally entangled state theory eqns. (\ref{eq:X_XX_two_photon_corr_theo}) and (\ref{eq:X_XX_two_photon_corr_model}) with and without taking into account the detector timing resolutions, respectively, see \cref{sec:prec_osc}. The equations are only scaled in the coincidence coordinate (i.e., their y-axis) to match the experimental data, otherwise there are no free parameters. Parameters $\FSS$ and $T_1^{\X}$ are determined independently, cf. \cref{sec:fss,sec:lifetime}, respectively. Using the method of time-dependent two-photon tomography, the two-photon density matrix as a function of $\tau$ is determined utilizing maximum-likelihood estimation \cite{James2001, Winik2017, Hopfmann2021}. The resulting entanglement negativities and selected density matrices are illustrated in \cref{fig:Ent_neg}. \\

\bibliography{Biblography}

\begin{thebibliography}{44}%
\makeatletter
\providecommand \@ifxundefined [1]{%
 \@ifx{#1\undefined}
}%
\providecommand \@ifnum [1]{%
 \ifnum #1\expandafter \@firstoftwo
 \else \expandafter \@secondoftwo
 \fi
}%
\providecommand \@ifx [1]{%
 \ifx #1\expandafter \@firstoftwo
 \else \expandafter \@secondoftwo
 \fi
}%
\providecommand \natexlab [1]{#1}%
\providecommand \enquote  [1]{``#1''}%
\providecommand \bibnamefont  [1]{#1}%
\providecommand \bibfnamefont [1]{#1}%
\providecommand \citenamefont [1]{#1}%
\providecommand \href@noop [0]{\@secondoftwo}%
\providecommand \href [0]{\begingroup \@sanitize@url \@href}%
\providecommand \@href[1]{\@@startlink{#1}\@@href}%
\providecommand \@@href[1]{\endgroup#1\@@endlink}%
\providecommand \@sanitize@url [0]{\catcode `\\12\catcode `\$12\catcode `\&12\catcode `\#12\catcode `\^12\catcode `\_12\catcode `\%12\relax}%
\providecommand \@@startlink[1]{}%
\providecommand \@@endlink[0]{}%
\providecommand \url  [0]{\begingroup\@sanitize@url \@url }%
\providecommand \@url [1]{\endgroup\@href {#1}{\urlprefix }}%
\providecommand \urlprefix  [0]{URL }%
\providecommand \Eprint [0]{\href }%
\providecommand \doibase [0]{http://dx.doi.org/}%
\providecommand \selectlanguage [0]{\@gobble}%
\providecommand \bibinfo  [0]{\@secondoftwo}%
\providecommand \bibfield  [0]{\@secondoftwo}%
\providecommand \translation [1]{[#1]}%
\providecommand \BibitemOpen [0]{}%
\providecommand \bibitemStop [0]{}%
\providecommand \bibitemNoStop [0]{.\EOS\space}%
\providecommand \EOS [0]{\spacefactor3000\relax}%
\providecommand \BibitemShut  [1]{\csname bibitem#1\endcsname}%
\let\auto@bib@innerbib\@empty
\bibitem [{\citenamefont {Kim}\ \emph {et~al.}(2008)\citenamefont {Kim}, \citenamefont {Economou}, \citenamefont {B\ifmmode~\u{a}\else \u{a}\fi{}descu}, \citenamefont {Scheibner}, \citenamefont {Bracker}, \citenamefont {Bashkansky}, \citenamefont {Reinecke},\ and\ \citenamefont {Gammon}}]{Kim2008}%
  \BibitemOpen
  \bibfield  {author} {\bibinfo {author} {\bibfnamefont {D.}~\bibnamefont {Kim}}, \bibinfo {author} {\bibfnamefont {S.~E.}\ \bibnamefont {Economou}}, \bibinfo {author} {\bibfnamefont {i.~m. c.~C.}\ \bibnamefont {B\ifmmode~\u{a}\else \u{a}\fi{}descu}}, \bibinfo {author} {\bibfnamefont {M.}~\bibnamefont {Scheibner}}, \bibinfo {author} {\bibfnamefont {A.~S.}\ \bibnamefont {Bracker}}, \bibinfo {author} {\bibfnamefont {M.}~\bibnamefont {Bashkansky}}, \bibinfo {author} {\bibfnamefont {T.~L.}\ \bibnamefont {Reinecke}}, \ and\ \bibinfo {author} {\bibfnamefont {D.}~\bibnamefont {Gammon}},\ }\bibfield  {title} {\enquote {\bibinfo {title} {Optical spin initialization and nondestructive measurement in a quantum dot molecule},}\ }\href {\doibase 10.1103/PhysRevLett.101.236804} {\bibfield  {journal} {\bibinfo  {journal} {Phys. Rev. Lett.}\ }\textbf {\bibinfo {volume} {101}},\ \bibinfo {pages} {236804} (\bibinfo {year} {2008})}\BibitemShut {NoStop}%
\bibitem [{\citenamefont {Wei}\ \emph {et~al.}(2022)\citenamefont {Wei}, \citenamefont {Jing}, \citenamefont {Zhang}, \citenamefont {Liao}, \citenamefont {Yuan}, \citenamefont {Fan}, \citenamefont {Lyu}, \citenamefont {Zhou}, \citenamefont {Wang}, \citenamefont {Deng}, \citenamefont {Song}, \citenamefont {Oblak}, \citenamefont {Guo},\ and\ \citenamefont {Zhou}}]{Wei2022}%
  \BibitemOpen
  \bibfield  {author} {\bibinfo {author} {\bibfnamefont {S.-H.}\ \bibnamefont {Wei}}, \bibinfo {author} {\bibfnamefont {B.}~\bibnamefont {Jing}}, \bibinfo {author} {\bibfnamefont {X.-Y.}\ \bibnamefont {Zhang}}, \bibinfo {author} {\bibfnamefont {J.-Y.}\ \bibnamefont {Liao}}, \bibinfo {author} {\bibfnamefont {C.-Z.}\ \bibnamefont {Yuan}}, \bibinfo {author} {\bibfnamefont {B.-Y.}\ \bibnamefont {Fan}}, \bibinfo {author} {\bibfnamefont {C.}~\bibnamefont {Lyu}}, \bibinfo {author} {\bibfnamefont {D.-L.}\ \bibnamefont {Zhou}}, \bibinfo {author} {\bibfnamefont {Y.}~\bibnamefont {Wang}}, \bibinfo {author} {\bibfnamefont {G.-W.}\ \bibnamefont {Deng}}, \bibinfo {author} {\bibfnamefont {H.-Z.}\ \bibnamefont {Song}}, \bibinfo {author} {\bibfnamefont {D.}~\bibnamefont {Oblak}}, \bibinfo {author} {\bibfnamefont {G.-C.}\ \bibnamefont {Guo}}, \ and\ \bibinfo {author} {\bibfnamefont {Q.}~\bibnamefont {Zhou}},\ }\bibfield  {title} {\enquote {\bibinfo {title} {Towards real-world quantum networks: A review},}\ }\href {\doibase
  https://doi.org/10.1002/lpor.202100219} {\bibfield  {journal} {\bibinfo  {journal} {Laser and Photonics Reviews}\ }\textbf {\bibinfo {volume} {16}},\ \bibinfo {pages} {2100219} (\bibinfo {year} {2022})}\BibitemShut {NoStop}%
\bibitem [{\citenamefont {Claudon}\ \emph {et~al.}(2010)\citenamefont {Claudon}, \citenamefont {Bleuse}, \citenamefont {Malik}, \citenamefont {Bazin}, \citenamefont {Jaffrennou}, \citenamefont {Gregersen}, \citenamefont {Sauvan}, \citenamefont {Lalanne},\ and\ \citenamefont {Gérard}}]{Claudon2010}%
  \BibitemOpen
  \bibfield  {author} {\bibinfo {author} {\bibfnamefont {J.}~\bibnamefont {Claudon}}, \bibinfo {author} {\bibfnamefont {J.}~\bibnamefont {Bleuse}}, \bibinfo {author} {\bibfnamefont {N.~S.}\ \bibnamefont {Malik}}, \bibinfo {author} {\bibfnamefont {M.}~\bibnamefont {Bazin}}, \bibinfo {author} {\bibfnamefont {P.}~\bibnamefont {Jaffrennou}}, \bibinfo {author} {\bibfnamefont {N.}~\bibnamefont {Gregersen}}, \bibinfo {author} {\bibfnamefont {C.}~\bibnamefont {Sauvan}}, \bibinfo {author} {\bibfnamefont {P.}~\bibnamefont {Lalanne}}, \ and\ \bibinfo {author} {\bibfnamefont {J.-M.}\ \bibnamefont {Gérard}},\ }\bibfield  {title} {\enquote {\bibinfo {title} {A highly efficient single-photon source based on a quantum dot in a photonic nanowire},}\ }\href {\doibase 10.1038/nphoton.2009.287x} {\bibfield  {journal} {\bibinfo  {journal} {Nature Photonics}\ }\textbf {\bibinfo {volume} {4}},\ \bibinfo {pages} {174--177} (\bibinfo {year} {2010})}\BibitemShut {NoStop}%
\bibitem [{\citenamefont {Dousse}\ \emph {et~al.}(2010)\citenamefont {Dousse}, \citenamefont {Suffczyński}, \citenamefont {Beveratos}, \citenamefont {Krebs}, \citenamefont {Lemaître}, \citenamefont {Sagnes}, \citenamefont {Bloch}, \citenamefont {Voisin},\ and\ \citenamefont {Senellart}}]{Dousse2010}%
  \BibitemOpen
  \bibfield  {author} {\bibinfo {author} {\bibfnamefont {A.}~\bibnamefont {Dousse}}, \bibinfo {author} {\bibfnamefont {J.}~\bibnamefont {Suffczyński}}, \bibinfo {author} {\bibfnamefont {A.}~\bibnamefont {Beveratos}}, \bibinfo {author} {\bibfnamefont {O.}~\bibnamefont {Krebs}}, \bibinfo {author} {\bibfnamefont {A.}~\bibnamefont {Lemaître}}, \bibinfo {author} {\bibfnamefont {I.}~\bibnamefont {Sagnes}}, \bibinfo {author} {\bibfnamefont {J.}~\bibnamefont {Bloch}}, \bibinfo {author} {\bibfnamefont {P.}~\bibnamefont {Voisin}}, \ and\ \bibinfo {author} {\bibfnamefont {P.}~\bibnamefont {Senellart}},\ }\bibfield  {title} {\enquote {\bibinfo {title} {Ultrabright source of entangled photon pairs},}\ }\href {\doibase 10.1038/nature09148} {\bibfield  {journal} {\bibinfo  {journal} {Nature}\ }\textbf {\bibinfo {volume} {466}},\ \bibinfo {pages} {217--220} (\bibinfo {year} {2010})}\BibitemShut {NoStop}%
\bibitem [{\citenamefont {Somaschi}\ \emph {et~al.}(2016)\citenamefont {Somaschi}, \citenamefont {Giesz}, \citenamefont {De~Santis}, \citenamefont {Loredo}, \citenamefont {Almeida}, \citenamefont {Hornecker}, \citenamefont {Portalupi}, \citenamefont {Grange}, \citenamefont {Antón}, \citenamefont {Demory}, \citenamefont {Gómez}, \citenamefont {Sagnes}, \citenamefont {Lanzillotti-Kimura}, \citenamefont {Lemaítre}, \citenamefont {Auffeves}, \citenamefont {White}, \citenamefont {Lanco},\ and\ \citenamefont {Senellart}}]{Somaschi2016}%
  \BibitemOpen
  \bibfield  {author} {\bibinfo {author} {\bibfnamefont {N.}~\bibnamefont {Somaschi}}, \bibinfo {author} {\bibfnamefont {V.}~\bibnamefont {Giesz}}, \bibinfo {author} {\bibfnamefont {L.}~\bibnamefont {De~Santis}}, \bibinfo {author} {\bibfnamefont {J.~C.}\ \bibnamefont {Loredo}}, \bibinfo {author} {\bibfnamefont {M.~P.}\ \bibnamefont {Almeida}}, \bibinfo {author} {\bibfnamefont {G.}~\bibnamefont {Hornecker}}, \bibinfo {author} {\bibfnamefont {S.~L.}\ \bibnamefont {Portalupi}}, \bibinfo {author} {\bibfnamefont {T.}~\bibnamefont {Grange}}, \bibinfo {author} {\bibfnamefont {C.}~\bibnamefont {Antón}}, \bibinfo {author} {\bibfnamefont {J.}~\bibnamefont {Demory}}, \bibinfo {author} {\bibfnamefont {C.}~\bibnamefont {Gómez}}, \bibinfo {author} {\bibfnamefont {I.}~\bibnamefont {Sagnes}}, \bibinfo {author} {\bibfnamefont {N.~D.}\ \bibnamefont {Lanzillotti-Kimura}}, \bibinfo {author} {\bibfnamefont {A.}~\bibnamefont {Lemaítre}}, \bibinfo {author} {\bibfnamefont {A.}~\bibnamefont {Auffeves}}, \bibinfo {author}
  {\bibfnamefont {A.~G.}\ \bibnamefont {White}}, \bibinfo {author} {\bibfnamefont {L.}~\bibnamefont {Lanco}}, \ and\ \bibinfo {author} {\bibfnamefont {P.}~\bibnamefont {Senellart}},\ }\bibfield  {title} {\enquote {\bibinfo {title} {Near-optimal single-photon sources in the solid state},}\ }\href {\doibase 10.1038/nphoton.2016.23} {\bibfield  {journal} {\bibinfo  {journal} {Nature Photonics}\ }\textbf {\bibinfo {volume} {10}},\ \bibinfo {pages} {340--345} (\bibinfo {year} {2016})}\BibitemShut {NoStop}%
\bibitem [{\citenamefont {Chen}\ \emph {et~al.}(2018)\citenamefont {Chen}, \citenamefont {Zopf}, \citenamefont {Keil}, \citenamefont {Ding},\ and\ \citenamefont {Schmidt}}]{Chen2018}%
  \BibitemOpen
  \bibfield  {author} {\bibinfo {author} {\bibfnamefont {Y.}~\bibnamefont {Chen}}, \bibinfo {author} {\bibfnamefont {M.}~\bibnamefont {Zopf}}, \bibinfo {author} {\bibfnamefont {R.}~\bibnamefont {Keil}}, \bibinfo {author} {\bibfnamefont {F.}~\bibnamefont {Ding}}, \ and\ \bibinfo {author} {\bibfnamefont {O.~G.}\ \bibnamefont {Schmidt}},\ }\bibfield  {title} {\enquote {\bibinfo {title} {Highly-efficient extraction of entangled photons from quantum dots using a broadband optical antenna},}\ }\href {\doibase 10.1038/s41467-018-05456-2} {\bibfield  {journal} {\bibinfo  {journal} {Nat. Commun.}\ }\textbf {\bibinfo {volume} {9}},\ \bibinfo {pages} {2994} (\bibinfo {year} {2018})}\BibitemShut {NoStop}%
\bibitem [{\citenamefont {Wang}\ \emph {et~al.}(2019)\citenamefont {Wang}, \citenamefont {Hu}, \citenamefont {Chung}, \citenamefont {Qin}, \citenamefont {Yang}, \citenamefont {Li}, \citenamefont {Liu}, \citenamefont {Zhong}, \citenamefont {He}, \citenamefont {Ding}, \citenamefont {Deng}, \citenamefont {Dai}, \citenamefont {Huo}, \citenamefont {H\"ofling}, \citenamefont {Lu},\ and\ \citenamefont {Pan}}]{Wang2019}%
  \BibitemOpen
  \bibfield  {author} {\bibinfo {author} {\bibfnamefont {H.}~\bibnamefont {Wang}}, \bibinfo {author} {\bibfnamefont {H.}~\bibnamefont {Hu}}, \bibinfo {author} {\bibfnamefont {T.-H.}\ \bibnamefont {Chung}}, \bibinfo {author} {\bibfnamefont {J.}~\bibnamefont {Qin}}, \bibinfo {author} {\bibfnamefont {X.}~\bibnamefont {Yang}}, \bibinfo {author} {\bibfnamefont {J.-P.}\ \bibnamefont {Li}}, \bibinfo {author} {\bibfnamefont {R.-Z.}\ \bibnamefont {Liu}}, \bibinfo {author} {\bibfnamefont {H.-S.}\ \bibnamefont {Zhong}}, \bibinfo {author} {\bibfnamefont {Y.-M.}\ \bibnamefont {He}}, \bibinfo {author} {\bibfnamefont {X.}~\bibnamefont {Ding}}, \bibinfo {author} {\bibfnamefont {Y.-H.}\ \bibnamefont {Deng}}, \bibinfo {author} {\bibfnamefont {Q.}~\bibnamefont {Dai}}, \bibinfo {author} {\bibfnamefont {Y.-H.}\ \bibnamefont {Huo}}, \bibinfo {author} {\bibfnamefont {S.}~\bibnamefont {H\"ofling}}, \bibinfo {author} {\bibfnamefont {C.-Y.}\ \bibnamefont {Lu}}, \ and\ \bibinfo {author} {\bibfnamefont {J.-W.}\ \bibnamefont {Pan}},\
  }\bibfield  {title} {\enquote {\bibinfo {title} {On-demand semiconductor source of entangled photons which simultaneously has high fidelity, efficiency, and indistinguishability},}\ }\href {\doibase 10.1103/PhysRevLett.122.113602} {\bibfield  {journal} {\bibinfo  {journal} {Phys. Rev. Lett.}\ }\textbf {\bibinfo {volume} {122}},\ \bibinfo {pages} {113602} (\bibinfo {year} {2019})}\BibitemShut {NoStop}%
\bibitem [{\citenamefont {Liu}\ \emph {et~al.}(2019)\citenamefont {Liu}, \citenamefont {Su}, \citenamefont {Wei}, \citenamefont {Yao}, \citenamefont {Silva}, \citenamefont {Yu}, \citenamefont {Iles-Smith}, \citenamefont {Srinivasan}, \citenamefont {Rastelli}, \citenamefont {Li},\ and\ \citenamefont {Wang}}]{Liu2019}%
  \BibitemOpen
  \bibfield  {author} {\bibinfo {author} {\bibfnamefont {J.}~\bibnamefont {Liu}}, \bibinfo {author} {\bibfnamefont {R.}~\bibnamefont {Su}}, \bibinfo {author} {\bibfnamefont {Y.}~\bibnamefont {Wei}}, \bibinfo {author} {\bibfnamefont {B.}~\bibnamefont {Yao}}, \bibinfo {author} {\bibfnamefont {S.~F. C.~d.}\ \bibnamefont {Silva}}, \bibinfo {author} {\bibfnamefont {Y.}~\bibnamefont {Yu}}, \bibinfo {author} {\bibfnamefont {J.}~\bibnamefont {Iles-Smith}}, \bibinfo {author} {\bibfnamefont {K.}~\bibnamefont {Srinivasan}}, \bibinfo {author} {\bibfnamefont {A.}~\bibnamefont {Rastelli}}, \bibinfo {author} {\bibfnamefont {J.}~\bibnamefont {Li}}, \ and\ \bibinfo {author} {\bibfnamefont {X.}~\bibnamefont {Wang}},\ }\bibfield  {title} {\enquote {\bibinfo {title} {A solid-state source of strongly entangled photon pairs with high brightness and indistinguishability},}\ }\href {\doibase 10.1038/s41565-019-0435-9} {\bibfield  {journal} {\bibinfo  {journal} {Nature Nanotechnology}\ }\textbf {\bibinfo {volume} {14}},\ \bibinfo
  {pages} {586--593} (\bibinfo {year} {2019})}\BibitemShut {NoStop}%
\bibitem [{\citenamefont {Lu}\ and\ \citenamefont {Pan}(2021)}]{Lu2021}%
  \BibitemOpen
  \bibfield  {author} {\bibinfo {author} {\bibfnamefont {C.-Y.}\ \bibnamefont {Lu}}\ and\ \bibinfo {author} {\bibfnamefont {J.-W.}\ \bibnamefont {Pan}},\ }\bibfield  {title} {\enquote {\bibinfo {title} {Quantum-dot single-photon sources for the quantum internet},}\ }\href {\doibase 10.1038/s41565-021-01033-9} {\bibfield  {journal} {\bibinfo  {journal} {Nature Nanotechnology}\ }\textbf {\bibinfo {volume} {16}},\ \bibinfo {pages} {1294--1296} (\bibinfo {year} {2021})}\BibitemShut {NoStop}%
\bibitem [{\citenamefont {Tomm}\ \emph {et~al.}(2021)\citenamefont {Tomm}, \citenamefont {Javadi}, \citenamefont {Antoniadis}, \citenamefont {Najer}, \citenamefont {Löbl}, \citenamefont {Korsch}, \citenamefont {Schott}, \citenamefont {Valentin}, \citenamefont {Wieck}, \citenamefont {Ludwig},\ and\ \citenamefont {Warburton}}]{Tomm2021}%
  \BibitemOpen
  \bibfield  {author} {\bibinfo {author} {\bibfnamefont {N.}~\bibnamefont {Tomm}}, \bibinfo {author} {\bibfnamefont {A.}~\bibnamefont {Javadi}}, \bibinfo {author} {\bibfnamefont {N.~O.}\ \bibnamefont {Antoniadis}}, \bibinfo {author} {\bibfnamefont {D.}~\bibnamefont {Najer}}, \bibinfo {author} {\bibfnamefont {M.~C.}\ \bibnamefont {Löbl}}, \bibinfo {author} {\bibfnamefont {A.~R.}\ \bibnamefont {Korsch}}, \bibinfo {author} {\bibfnamefont {R.}~\bibnamefont {Schott}}, \bibinfo {author} {\bibfnamefont {S.~R.}\ \bibnamefont {Valentin}}, \bibinfo {author} {\bibfnamefont {A.~D.}\ \bibnamefont {Wieck}}, \bibinfo {author} {\bibfnamefont {A.}~\bibnamefont {Ludwig}}, \ and\ \bibinfo {author} {\bibfnamefont {R.~J.}\ \bibnamefont {Warburton}},\ }\bibfield  {title} {\enquote {\bibinfo {title} {A bright and fast source of coherent single photons},}\ }\href {\doibase 10.1038/s41565-020-00831-x} {\bibfield  {journal} {\bibinfo  {journal} {Nature Nanotechnology}\ }\textbf {\bibinfo {volume} {16}},\ \bibinfo {pages} {399--403}
  (\bibinfo {year} {2021})}\BibitemShut {NoStop}%
\bibitem [{\citenamefont {Shooter}\ \emph {et~al.}(2020)\citenamefont {Shooter}, \citenamefont {Xiang}, \citenamefont {M\"{u}ller}, \citenamefont {Skiba-Szymanska}, \citenamefont {Huwer}, \citenamefont {Griffiths}, \citenamefont {Mitchell}, \citenamefont {Anderson}, \citenamefont {M\"{u}ller}, \citenamefont {Krysa}, \citenamefont {Stevenson}, \citenamefont {Heffernan}, \citenamefont {Ritchie},\ and\ \citenamefont {Shields}}]{Shooter2020}%
  \BibitemOpen
  \bibfield  {author} {\bibinfo {author} {\bibfnamefont {G.}~\bibnamefont {Shooter}}, \bibinfo {author} {\bibfnamefont {Z.-H.}\ \bibnamefont {Xiang}}, \bibinfo {author} {\bibfnamefont {J.~R.~A.}\ \bibnamefont {M\"{u}ller}}, \bibinfo {author} {\bibfnamefont {J.}~\bibnamefont {Skiba-Szymanska}}, \bibinfo {author} {\bibfnamefont {J.}~\bibnamefont {Huwer}}, \bibinfo {author} {\bibfnamefont {J.}~\bibnamefont {Griffiths}}, \bibinfo {author} {\bibfnamefont {T.}~\bibnamefont {Mitchell}}, \bibinfo {author} {\bibfnamefont {M.}~\bibnamefont {Anderson}}, \bibinfo {author} {\bibfnamefont {T.}~\bibnamefont {M\"{u}ller}}, \bibinfo {author} {\bibfnamefont {A.~B.}\ \bibnamefont {Krysa}}, \bibinfo {author} {\bibfnamefont {R.~M.}\ \bibnamefont {Stevenson}}, \bibinfo {author} {\bibfnamefont {J.}~\bibnamefont {Heffernan}}, \bibinfo {author} {\bibfnamefont {D.~A.}\ \bibnamefont {Ritchie}}, \ and\ \bibinfo {author} {\bibfnamefont {A.~J.}\ \bibnamefont {Shields}},\ }\bibfield  {title} {\enquote {\bibinfo {title} {1{GHz} clocked
  distribution of electrically generated entangled photon pairs},}\ }\href {\doibase 10.1364/OE.405466} {\bibfield  {journal} {\bibinfo  {journal} {Opt. Express}\ }\textbf {\bibinfo {volume} {28}},\ \bibinfo {pages} {36838--36848} (\bibinfo {year} {2020})}\BibitemShut {NoStop}%
\bibitem [{\citenamefont {Uppu}\ \emph {et~al.}(2020)\citenamefont {Uppu}, \citenamefont {Pedersen}, \citenamefont {Wang}, \citenamefont {Olesen}, \citenamefont {Papon}, \citenamefont {Zhou}, \citenamefont {Midolo}, \citenamefont {Scholz}, \citenamefont {Wieck}, \citenamefont {Ludwig},\ and\ \citenamefont {Lodahl}}]{Uppu2020}%
  \BibitemOpen
  \bibfield  {author} {\bibinfo {author} {\bibfnamefont {R.}~\bibnamefont {Uppu}}, \bibinfo {author} {\bibfnamefont {F.~T.}\ \bibnamefont {Pedersen}}, \bibinfo {author} {\bibfnamefont {Y.}~\bibnamefont {Wang}}, \bibinfo {author} {\bibfnamefont {C.~T.}\ \bibnamefont {Olesen}}, \bibinfo {author} {\bibfnamefont {C.}~\bibnamefont {Papon}}, \bibinfo {author} {\bibfnamefont {X.}~\bibnamefont {Zhou}}, \bibinfo {author} {\bibfnamefont {L.}~\bibnamefont {Midolo}}, \bibinfo {author} {\bibfnamefont {S.}~\bibnamefont {Scholz}}, \bibinfo {author} {\bibfnamefont {A.~D.}\ \bibnamefont {Wieck}}, \bibinfo {author} {\bibfnamefont {A.}~\bibnamefont {Ludwig}}, \ and\ \bibinfo {author} {\bibfnamefont {P.}~\bibnamefont {Lodahl}},\ }\bibfield  {title} {\enquote {\bibinfo {title} {Scalable integrated single-photon source},}\ }\href {\doibase 10.1126/sciadv.abc8268} {\bibfield  {journal} {\bibinfo  {journal} {Science Advances}\ }\textbf {\bibinfo {volume} {6}},\ \bibinfo {pages} {eabc8268} (\bibinfo {year} {2020})}\BibitemShut
  {NoStop}%
\bibitem [{\citenamefont {Hopfmann}\ \emph {et~al.}(2021{\natexlab{a}})\citenamefont {Hopfmann}, \citenamefont {Nie}, \citenamefont {Sharma}, \citenamefont {Weigelt}, \citenamefont {Ding},\ and\ \citenamefont {Schmidt}}]{Hopfmann2021}%
  \BibitemOpen
  \bibfield  {author} {\bibinfo {author} {\bibfnamefont {C.}~\bibnamefont {Hopfmann}}, \bibinfo {author} {\bibfnamefont {W.}~\bibnamefont {Nie}}, \bibinfo {author} {\bibfnamefont {N.~L.}\ \bibnamefont {Sharma}}, \bibinfo {author} {\bibfnamefont {C.}~\bibnamefont {Weigelt}}, \bibinfo {author} {\bibfnamefont {F.}~\bibnamefont {Ding}}, \ and\ \bibinfo {author} {\bibfnamefont {O.~G.}\ \bibnamefont {Schmidt}},\ }\bibfield  {title} {\enquote {\bibinfo {title} {Maximally entangled and gigahertz-clocked on-demand photon pair source},}\ }\href {\doibase 10.1103/PhysRevB.103.075413} {\bibfield  {journal} {\bibinfo  {journal} {Phys. Rev. B}\ }\textbf {\bibinfo {volume} {103}},\ \bibinfo {pages} {075413} (\bibinfo {year} {2021}{\natexlab{a}})}\BibitemShut {NoStop}%
\bibitem [{\citenamefont {van Loock}\ \emph {et~al.}(2020)\citenamefont {van Loock}, \citenamefont {Alt}, \citenamefont {Becher}, \citenamefont {Benson}, \citenamefont {Boche}, \citenamefont {Deppe}, \citenamefont {Eschner}, \citenamefont {Höfling}, \citenamefont {Meschede}, \citenamefont {Michler}, \citenamefont {Schmidt},\ and\ \citenamefont {Weinfurter}}]{Loock2020}%
  \BibitemOpen
  \bibfield  {author} {\bibinfo {author} {\bibfnamefont {P.}~\bibnamefont {van Loock}}, \bibinfo {author} {\bibfnamefont {W.}~\bibnamefont {Alt}}, \bibinfo {author} {\bibfnamefont {C.}~\bibnamefont {Becher}}, \bibinfo {author} {\bibfnamefont {O.}~\bibnamefont {Benson}}, \bibinfo {author} {\bibfnamefont {H.}~\bibnamefont {Boche}}, \bibinfo {author} {\bibfnamefont {C.}~\bibnamefont {Deppe}}, \bibinfo {author} {\bibfnamefont {J.}~\bibnamefont {Eschner}}, \bibinfo {author} {\bibfnamefont {S.}~\bibnamefont {Höfling}}, \bibinfo {author} {\bibfnamefont {D.}~\bibnamefont {Meschede}}, \bibinfo {author} {\bibfnamefont {P.}~\bibnamefont {Michler}}, \bibinfo {author} {\bibfnamefont {F.}~\bibnamefont {Schmidt}}, \ and\ \bibinfo {author} {\bibfnamefont {H.}~\bibnamefont {Weinfurter}},\ }\bibfield  {title} {\enquote {\bibinfo {title} {Extending {Q}uantum {L}inks: {M}odules for {F}iber- and {M}emory-{B}ased {Q}uantum {R}epeaters},}\ }\href {\doibase https://doi.org/10.1002/qute.201900141} {\bibfield  {journal} {\bibinfo
  {journal} {Advanced Quantum Technologies}\ }\textbf {\bibinfo {volume} {3}},\ \bibinfo {pages} {1900141} (\bibinfo {year} {2020})}\BibitemShut {NoStop}%
\bibitem [{\citenamefont {Schimpf}\ \emph {et~al.}(2021)\citenamefont {Schimpf}, \citenamefont {Reindl}, \citenamefont {Basso~Basset}, \citenamefont {Jöns}, \citenamefont {Trotta},\ and\ \citenamefont {Rastelli}}]{Schimpf2021}%
  \BibitemOpen
  \bibfield  {author} {\bibinfo {author} {\bibfnamefont {C.}~\bibnamefont {Schimpf}}, \bibinfo {author} {\bibfnamefont {M.}~\bibnamefont {Reindl}}, \bibinfo {author} {\bibfnamefont {F.}~\bibnamefont {Basso~Basset}}, \bibinfo {author} {\bibfnamefont {K.~D.}\ \bibnamefont {Jöns}}, \bibinfo {author} {\bibfnamefont {R.}~\bibnamefont {Trotta}}, \ and\ \bibinfo {author} {\bibfnamefont {A.}~\bibnamefont {Rastelli}},\ }\bibfield  {title} {\enquote {\bibinfo {title} {{Quantum dots as potential sources of strongly entangled photons: Perspectives and challenges for applications in quantum networks}},}\ }\href {\doibase 10.1063/5.0038729} {\bibfield  {journal} {\bibinfo  {journal} {Applied Physics Letters}\ }\textbf {\bibinfo {volume} {118}},\ \bibinfo {pages} {100502} (\bibinfo {year} {2021})}\BibitemShut {NoStop}%
\bibitem [{\citenamefont {Wang}\ \emph {et~al.}(2016)\citenamefont {Wang}, \citenamefont {Chen}, \citenamefont {Li}, \citenamefont {Huang}, \citenamefont {Liu}, \citenamefont {Chen}, \citenamefont {Luo}, \citenamefont {Su}, \citenamefont {Wu}, \citenamefont {Li}, \citenamefont {Lu}, \citenamefont {Hu}, \citenamefont {Jiang}, \citenamefont {Peng}, \citenamefont {Li}, \citenamefont {Liu}, \citenamefont {Chen}, \citenamefont {Lu},\ and\ \citenamefont {Pan}}]{Wang2016}%
  \BibitemOpen
  \bibfield  {author} {\bibinfo {author} {\bibfnamefont {X.-L.}\ \bibnamefont {Wang}}, \bibinfo {author} {\bibfnamefont {L.-K.}\ \bibnamefont {Chen}}, \bibinfo {author} {\bibfnamefont {W.}~\bibnamefont {Li}}, \bibinfo {author} {\bibfnamefont {H.-L.}\ \bibnamefont {Huang}}, \bibinfo {author} {\bibfnamefont {C.}~\bibnamefont {Liu}}, \bibinfo {author} {\bibfnamefont {C.}~\bibnamefont {Chen}}, \bibinfo {author} {\bibfnamefont {Y.-H.}\ \bibnamefont {Luo}}, \bibinfo {author} {\bibfnamefont {Z.-E.}\ \bibnamefont {Su}}, \bibinfo {author} {\bibfnamefont {D.}~\bibnamefont {Wu}}, \bibinfo {author} {\bibfnamefont {Z.-D.}\ \bibnamefont {Li}}, \bibinfo {author} {\bibfnamefont {H.}~\bibnamefont {Lu}}, \bibinfo {author} {\bibfnamefont {Y.}~\bibnamefont {Hu}}, \bibinfo {author} {\bibfnamefont {X.}~\bibnamefont {Jiang}}, \bibinfo {author} {\bibfnamefont {C.-Z.}\ \bibnamefont {Peng}}, \bibinfo {author} {\bibfnamefont {L.}~\bibnamefont {Li}}, \bibinfo {author} {\bibfnamefont {N.-L.}\ \bibnamefont {Liu}}, \bibinfo {author}
  {\bibfnamefont {Y.-A.}\ \bibnamefont {Chen}}, \bibinfo {author} {\bibfnamefont {C.-Y.}\ \bibnamefont {Lu}}, \ and\ \bibinfo {author} {\bibfnamefont {J.-W.}\ \bibnamefont {Pan}},\ }\bibfield  {title} {\enquote {\bibinfo {title} {Experimental ten-photon entanglement},}\ }\href {\doibase 10.1103/PhysRevLett.117.210502} {\bibfield  {journal} {\bibinfo  {journal} {Phys. Rev. Lett.}\ }\textbf {\bibinfo {volume} {117}},\ \bibinfo {pages} {210502} (\bibinfo {year} {2016})}\BibitemShut {NoStop}%
\bibitem [{\citenamefont {McCutcheon}\ \emph {et~al.}(2016)\citenamefont {McCutcheon}, \citenamefont {Pappa}, \citenamefont {Bell}, \citenamefont {McMillan}, \citenamefont {Chailloux}, \citenamefont {Lawson}, \citenamefont {Mafu}, \citenamefont {Markham}, \citenamefont {Diamanti}, \citenamefont {Kerenidis}, \citenamefont {Rarity},\ and\ \citenamefont {Tame}}]{McCutcheon2016}%
  \BibitemOpen
  \bibfield  {author} {\bibinfo {author} {\bibfnamefont {W.}~\bibnamefont {McCutcheon}}, \bibinfo {author} {\bibfnamefont {A.}~\bibnamefont {Pappa}}, \bibinfo {author} {\bibfnamefont {B.~A.}\ \bibnamefont {Bell}}, \bibinfo {author} {\bibfnamefont {A.}~\bibnamefont {McMillan}}, \bibinfo {author} {\bibfnamefont {A.}~\bibnamefont {Chailloux}}, \bibinfo {author} {\bibfnamefont {T.}~\bibnamefont {Lawson}}, \bibinfo {author} {\bibfnamefont {M.}~\bibnamefont {Mafu}}, \bibinfo {author} {\bibfnamefont {D.}~\bibnamefont {Markham}}, \bibinfo {author} {\bibfnamefont {E.}~\bibnamefont {Diamanti}}, \bibinfo {author} {\bibfnamefont {I.}~\bibnamefont {Kerenidis}}, \bibinfo {author} {\bibfnamefont {J.~G.}\ \bibnamefont {Rarity}}, \ and\ \bibinfo {author} {\bibfnamefont {M.~S.}\ \bibnamefont {Tame}},\ }\bibfield  {title} {\enquote {\bibinfo {title} {Experimental verification of multipartite entanglement in quantum networks},}\ }\href {\doibase 10.1038/ncomms13251} {\bibfield  {journal} {\bibinfo  {journal} {Nature
  Communications}\ }\textbf {\bibinfo {volume} {7}},\ \bibinfo {pages} {13251} (\bibinfo {year} {2016})}\BibitemShut {NoStop}%
\bibitem [{\citenamefont {Zhang}\ \emph {et~al.}(2021)\citenamefont {Zhang}, \citenamefont {Huang}, \citenamefont {Liu}, \citenamefont {Li},\ and\ \citenamefont {Guo}}]{Zhang2021a}%
  \BibitemOpen
  \bibfield  {author} {\bibinfo {author} {\bibfnamefont {C.}~\bibnamefont {Zhang}}, \bibinfo {author} {\bibfnamefont {Y.-F.}\ \bibnamefont {Huang}}, \bibinfo {author} {\bibfnamefont {B.-H.}\ \bibnamefont {Liu}}, \bibinfo {author} {\bibfnamefont {C.-F.}\ \bibnamefont {Li}}, \ and\ \bibinfo {author} {\bibfnamefont {G.-C.}\ \bibnamefont {Guo}},\ }\bibfield  {title} {\enquote {\bibinfo {title} {Spontaneous parametric down-conversion sources for multiphoton experiments},}\ }\href {\doibase https://doi.org/10.1002/qute.202000132} {\bibfield  {journal} {\bibinfo  {journal} {Advanced Quantum Technologies}\ }\textbf {\bibinfo {volume} {4}},\ \bibinfo {pages} {2000132} (\bibinfo {year} {2021})}\BibitemShut {NoStop}%
\bibitem [{\citenamefont {Cacciapuoti}\ \emph {et~al.}(2020)\citenamefont {Cacciapuoti}, \citenamefont {Caleffi}, \citenamefont {Tafuri}, \citenamefont {Cataliotti}, \citenamefont {Gherardini},\ and\ \citenamefont {Bianchi}}]{Cacciapuoti2020}%
  \BibitemOpen
  \bibfield  {author} {\bibinfo {author} {\bibfnamefont {A.~S.}\ \bibnamefont {Cacciapuoti}}, \bibinfo {author} {\bibfnamefont {M.}~\bibnamefont {Caleffi}}, \bibinfo {author} {\bibfnamefont {F.}~\bibnamefont {Tafuri}}, \bibinfo {author} {\bibfnamefont {F.~S.}\ \bibnamefont {Cataliotti}}, \bibinfo {author} {\bibfnamefont {S.}~\bibnamefont {Gherardini}}, \ and\ \bibinfo {author} {\bibfnamefont {G.}~\bibnamefont {Bianchi}},\ }\bibfield  {title} {\enquote {\bibinfo {title} {Quantum internet: Networking challenges in distributed quantum computing},}\ }\href {\doibase 10.1109/mnet.001.1900092} {\bibfield  {journal} {\bibinfo  {journal} {{IEEE} Network}\ }\textbf {\bibinfo {volume} {34}},\ \bibinfo {pages} {137--143} (\bibinfo {year} {2020})}\BibitemShut {NoStop}%
\bibitem [{\citenamefont {Basset}\ \emph {et~al.}(2021)\citenamefont {Basset}, \citenamefont {Valeri}, \citenamefont {Roccia}, \citenamefont {Muredda}, \citenamefont {Poderini}, \citenamefont {Neuwirth}, \citenamefont {Spagnolo}, \citenamefont {Rota}, \citenamefont {Carvacho}, \citenamefont {Sciarrino},\ and\ \citenamefont {Trotta}}]{Basset2021}%
  \BibitemOpen
  \bibfield  {author} {\bibinfo {author} {\bibfnamefont {F.~B.}\ \bibnamefont {Basset}}, \bibinfo {author} {\bibfnamefont {M.}~\bibnamefont {Valeri}}, \bibinfo {author} {\bibfnamefont {E.}~\bibnamefont {Roccia}}, \bibinfo {author} {\bibfnamefont {V.}~\bibnamefont {Muredda}}, \bibinfo {author} {\bibfnamefont {D.}~\bibnamefont {Poderini}}, \bibinfo {author} {\bibfnamefont {J.}~\bibnamefont {Neuwirth}}, \bibinfo {author} {\bibfnamefont {N.}~\bibnamefont {Spagnolo}}, \bibinfo {author} {\bibfnamefont {M.~B.}\ \bibnamefont {Rota}}, \bibinfo {author} {\bibfnamefont {G.}~\bibnamefont {Carvacho}}, \bibinfo {author} {\bibfnamefont {F.}~\bibnamefont {Sciarrino}}, \ and\ \bibinfo {author} {\bibfnamefont {R.}~\bibnamefont {Trotta}},\ }\bibfield  {title} {\enquote {\bibinfo {title} {Quantum key distribution with entangled photons generated on demand by a quantum dot},}\ }\href {\doibase 10.1126/sciadv.abe6379} {\bibfield  {journal} {\bibinfo  {journal} {Science Advances}\ }\textbf {\bibinfo {volume} {7}},\ \bibinfo {pages}
  {eabe6379} (\bibinfo {year} {2021})}\BibitemShut {NoStop}%
\bibitem [{\citenamefont {Stevens}(2013)}]{Stevens2013}%
  \BibitemOpen
  \bibfield  {author} {\bibinfo {author} {\bibfnamefont {M.~J.}\ \bibnamefont {Stevens}},\ }\bibfield  {title} {\enquote {\bibinfo {title} {Chapter 2 - photon statistics, measurements, and measurements tools},}\ }in\ \href {\doibase https://doi.org/10.1016/B978-0-12-387695-9.00002-0} {\emph {\bibinfo {booktitle} {Single-Photon Generation and Detection}}},\ \bibinfo {series} {Experimental Methods in the Physical Sciences}, Vol.~\bibinfo {volume} {45},\ \bibinfo {editor} {edited by\ \bibinfo {editor} {\bibfnamefont {A.}~\bibnamefont {Migdall}}, \bibinfo {editor} {\bibfnamefont {S.~V.}\ \bibnamefont {Polyakov}}, \bibinfo {editor} {\bibfnamefont {J.}~\bibnamefont {Fan}}, \ and\ \bibinfo {editor} {\bibfnamefont {J.~C.}\ \bibnamefont {Bienfang}}}\ (\bibinfo  {publisher} {Academic Press},\ \bibinfo {year} {2013})\ pp.\ \bibinfo {pages} {25--68}\BibitemShut {NoStop}%
\bibitem [{\citenamefont {Schlehahn}\ \emph {et~al.}(2018)\citenamefont {Schlehahn}, \citenamefont {Fischbach}, \citenamefont {Schmidt}, \citenamefont {Kaganskiy}, \citenamefont {Strittmatter}, \citenamefont {Rodt}, \citenamefont {Heindel},\ and\ \citenamefont {Reitzenstein}}]{Schlehahn2018}%
  \BibitemOpen
  \bibfield  {author} {\bibinfo {author} {\bibfnamefont {A.}~\bibnamefont {Schlehahn}}, \bibinfo {author} {\bibfnamefont {S.}~\bibnamefont {Fischbach}}, \bibinfo {author} {\bibfnamefont {R.}~\bibnamefont {Schmidt}}, \bibinfo {author} {\bibfnamefont {A.}~\bibnamefont {Kaganskiy}}, \bibinfo {author} {\bibfnamefont {A.}~\bibnamefont {Strittmatter}}, \bibinfo {author} {\bibfnamefont {S.}~\bibnamefont {Rodt}}, \bibinfo {author} {\bibfnamefont {T.}~\bibnamefont {Heindel}}, \ and\ \bibinfo {author} {\bibfnamefont {S.}~\bibnamefont {Reitzenstein}},\ }\bibfield  {title} {\enquote {\bibinfo {title} {A stand-alone fiber-coupled single-photon source},}\ }\href {\doibase 10.1038/s41598-017-19049-4} {\bibfield  {journal} {\bibinfo  {journal} {Scientific Reports}\ }\textbf {\bibinfo {volume} {8}},\ \bibinfo {pages} {1340} (\bibinfo {year} {2018})}\BibitemShut {NoStop}%
\bibitem [{\citenamefont {Musiał}\ \emph {et~al.}(2020)\citenamefont {Musiał}, \citenamefont {Żołnacz}, \citenamefont {Srocka}, \citenamefont {Kravets}, \citenamefont {Große}, \citenamefont {Olszewski}, \citenamefont {Poturaj}, \citenamefont {Wójcik}, \citenamefont {Mergo}, \citenamefont {Dybka}, \citenamefont {Dyrkacz}, \citenamefont {Dłubek}, \citenamefont {Lauritsen}, \citenamefont {Bülter}, \citenamefont {Schneider}, \citenamefont {Zschiedrich}, \citenamefont {Burger}, \citenamefont {Rodt}, \citenamefont {Urbańczyk}, \citenamefont {Sęk},\ and\ \citenamefont {Reitzenstein}}]{Musial2020}%
  \BibitemOpen
  \bibfield  {author} {\bibinfo {author} {\bibfnamefont {A.}~\bibnamefont {Musiał}}, \bibinfo {author} {\bibfnamefont {K.}~\bibnamefont {Żołnacz}}, \bibinfo {author} {\bibfnamefont {N.}~\bibnamefont {Srocka}}, \bibinfo {author} {\bibfnamefont {O.}~\bibnamefont {Kravets}}, \bibinfo {author} {\bibfnamefont {J.}~\bibnamefont {Große}}, \bibinfo {author} {\bibfnamefont {J.}~\bibnamefont {Olszewski}}, \bibinfo {author} {\bibfnamefont {K.}~\bibnamefont {Poturaj}}, \bibinfo {author} {\bibfnamefont {G.}~\bibnamefont {Wójcik}}, \bibinfo {author} {\bibfnamefont {P.}~\bibnamefont {Mergo}}, \bibinfo {author} {\bibfnamefont {K.}~\bibnamefont {Dybka}}, \bibinfo {author} {\bibfnamefont {M.}~\bibnamefont {Dyrkacz}}, \bibinfo {author} {\bibfnamefont {M.}~\bibnamefont {Dłubek}}, \bibinfo {author} {\bibfnamefont {K.}~\bibnamefont {Lauritsen}}, \bibinfo {author} {\bibfnamefont {A.}~\bibnamefont {Bülter}}, \bibinfo {author} {\bibfnamefont {P.-I.}\ \bibnamefont {Schneider}}, \bibinfo {author} {\bibfnamefont {L.}~\bibnamefont
  {Zschiedrich}}, \bibinfo {author} {\bibfnamefont {S.}~\bibnamefont {Burger}}, \bibinfo {author} {\bibfnamefont {S.}~\bibnamefont {Rodt}}, \bibinfo {author} {\bibfnamefont {W.}~\bibnamefont {Urbańczyk}}, \bibinfo {author} {\bibfnamefont {G.}~\bibnamefont {Sęk}}, \ and\ \bibinfo {author} {\bibfnamefont {S.}~\bibnamefont {Reitzenstein}},\ }\bibfield  {title} {\enquote {\bibinfo {title} {Plug\&play fiber-coupled 73 {kHz} single-photon source operating in the telecom o-band},}\ }\href {\doibase https://doi.org/10.1002/qute.202000018} {\bibfield  {journal} {\bibinfo  {journal} {Advanced Quantum Technologies}\ }\textbf {\bibinfo {volume} {3}},\ \bibinfo {pages} {2000018} (\bibinfo {year} {2020})}\BibitemShut {NoStop}%
\bibitem [{\citenamefont {Bremer}\ \emph {et~al.}(2020)\citenamefont {Bremer}, \citenamefont {Weber}, \citenamefont {Fischbach}, \citenamefont {Thiele}, \citenamefont {Schmidt}, \citenamefont {Kaganskiy}, \citenamefont {Rodt}, \citenamefont {Herkommer}, \citenamefont {Sartison}, \citenamefont {Portalupi}, \citenamefont {Michler}, \citenamefont {Giessen},\ and\ \citenamefont {Reitzenstein}}]{Bremer2020}%
  \BibitemOpen
  \bibfield  {author} {\bibinfo {author} {\bibfnamefont {L.}~\bibnamefont {Bremer}}, \bibinfo {author} {\bibfnamefont {K.}~\bibnamefont {Weber}}, \bibinfo {author} {\bibfnamefont {S.}~\bibnamefont {Fischbach}}, \bibinfo {author} {\bibfnamefont {S.}~\bibnamefont {Thiele}}, \bibinfo {author} {\bibfnamefont {M.}~\bibnamefont {Schmidt}}, \bibinfo {author} {\bibfnamefont {A.}~\bibnamefont {Kaganskiy}}, \bibinfo {author} {\bibfnamefont {S.}~\bibnamefont {Rodt}}, \bibinfo {author} {\bibfnamefont {A.}~\bibnamefont {Herkommer}}, \bibinfo {author} {\bibfnamefont {M.}~\bibnamefont {Sartison}}, \bibinfo {author} {\bibfnamefont {S.~L.}\ \bibnamefont {Portalupi}}, \bibinfo {author} {\bibfnamefont {P.}~\bibnamefont {Michler}}, \bibinfo {author} {\bibfnamefont {H.}~\bibnamefont {Giessen}}, \ and\ \bibinfo {author} {\bibfnamefont {S.}~\bibnamefont {Reitzenstein}},\ }\bibfield  {title} {\enquote {\bibinfo {title} {{Quantum dot single-photon emission coupled into single-mode fibers with 3D printed micro-objectives}},}\ }\href
  {\doibase 10.1063/5.0014921} {\bibfield  {journal} {\bibinfo  {journal} {APL Photonics}\ }\textbf {\bibinfo {volume} {5}},\ \bibinfo {pages} {106101} (\bibinfo {year} {2020})}\BibitemShut {NoStop}%
\bibitem [{\citenamefont {Aspnes}\ \emph {et~al.}(1986)\citenamefont {Aspnes}, \citenamefont {Kelso}, \citenamefont {Logan},\ and\ \citenamefont {Bhat}}]{Aspnes1986}%
  \BibitemOpen
  \bibfield  {author} {\bibinfo {author} {\bibfnamefont {D.~E.}\ \bibnamefont {Aspnes}}, \bibinfo {author} {\bibfnamefont {S.~M.}\ \bibnamefont {Kelso}}, \bibinfo {author} {\bibfnamefont {R.~A.}\ \bibnamefont {Logan}}, \ and\ \bibinfo {author} {\bibfnamefont {R.}~\bibnamefont {Bhat}},\ }\bibfield  {title} {\enquote {\bibinfo {title} {{Optical properties of AlxGa1-x As}},}\ }\href {\doibase 10.1063/1.337426} {\bibfield  {journal} {\bibinfo  {journal} {Journal of Applied Physics}\ }\textbf {\bibinfo {volume} {60}},\ \bibinfo {pages} {754--767} (\bibinfo {year} {1986})}\BibitemShut {NoStop}%
\bibitem [{\citenamefont {Shields}(2007)}]{Shields2007}%
  \BibitemOpen
  \bibfield  {author} {\bibinfo {author} {\bibfnamefont {A.~J.}\ \bibnamefont {Shields}},\ }\bibfield  {title} {\enquote {\bibinfo {title} {Semiconductor quantum light sources},}\ }\href {\doibase 10.1038/nphoton.2007.46} {\bibfield  {journal} {\bibinfo  {journal} {Nature Photonics}\ }\textbf {\bibinfo {volume} {1}},\ \bibinfo {pages} {215--223} (\bibinfo {year} {2007})}\BibitemShut {NoStop}%
\bibitem [{\citenamefont {Nie}\ \emph {et~al.}(2021)\citenamefont {Nie}, \citenamefont {Sharma}, \citenamefont {Weigelt}, \citenamefont {Keil}, \citenamefont {Yang}, \citenamefont {Ding}, \citenamefont {Hopfmann},\ and\ \citenamefont {Schmidt}}]{Nie2021}%
  \BibitemOpen
  \bibfield  {author} {\bibinfo {author} {\bibfnamefont {W.}~\bibnamefont {Nie}}, \bibinfo {author} {\bibfnamefont {N.~L.}\ \bibnamefont {Sharma}}, \bibinfo {author} {\bibfnamefont {C.}~\bibnamefont {Weigelt}}, \bibinfo {author} {\bibfnamefont {R.}~\bibnamefont {Keil}}, \bibinfo {author} {\bibfnamefont {J.}~\bibnamefont {Yang}}, \bibinfo {author} {\bibfnamefont {F.}~\bibnamefont {Ding}}, \bibinfo {author} {\bibfnamefont {C.}~\bibnamefont {Hopfmann}}, \ and\ \bibinfo {author} {\bibfnamefont {O.~G.}\ \bibnamefont {Schmidt}},\ }\bibfield  {title} {\enquote {\bibinfo {title} {Experimental optimization of the fiber coupling efficiency of {GaAs} quantum dot-based photon sources},}\ }\href {\doibase 10.1063/5.0059310} {\bibfield  {journal} {\bibinfo  {journal} {Applied Physics Letters}\ }\textbf {\bibinfo {volume} {119}},\ \bibinfo {pages} {244003} (\bibinfo {year} {2021})}\BibitemShut {NoStop}%
\bibitem [{\citenamefont {Stufler}\ \emph {et~al.}(2006)\citenamefont {Stufler}, \citenamefont {Machnikowski}, \citenamefont {Ester}, \citenamefont {Bichler}, \citenamefont {Axt}, \citenamefont {Kuhn},\ and\ \citenamefont {Zrenner}}]{Stufler2006}%
  \BibitemOpen
  \bibfield  {author} {\bibinfo {author} {\bibfnamefont {S.}~\bibnamefont {Stufler}}, \bibinfo {author} {\bibfnamefont {P.}~\bibnamefont {Machnikowski}}, \bibinfo {author} {\bibfnamefont {P.}~\bibnamefont {Ester}}, \bibinfo {author} {\bibfnamefont {M.}~\bibnamefont {Bichler}}, \bibinfo {author} {\bibfnamefont {V.~M.}\ \bibnamefont {Axt}}, \bibinfo {author} {\bibfnamefont {T.}~\bibnamefont {Kuhn}}, \ and\ \bibinfo {author} {\bibfnamefont {A.}~\bibnamefont {Zrenner}},\ }\bibfield  {title} {\enquote {\bibinfo {title} {{Two-photon Rabi oscillations in a single ${\mathrm{{I}n}}_{x}{\mathrm{{G}a}}_{1-x}\mathrm{{A}s}-\mathrm{{G}a}\mathrm{{A}s}$ quantum dot}},}\ }\href {\doibase 10.1103/PhysRevB.73.125304} {\bibfield  {journal} {\bibinfo  {journal} {Phys. Rev. B}\ }\textbf {\bibinfo {volume} {73}},\ \bibinfo {pages} {125304} (\bibinfo {year} {2006})}\BibitemShut {NoStop}%
\bibitem [{\citenamefont {Bounouar}\ \emph {et~al.}(2015)\citenamefont {Bounouar}, \citenamefont {M\"uller}, \citenamefont {Barth}, \citenamefont {Gl\"assl}, \citenamefont {Axt},\ and\ \citenamefont {Michler}}]{Bounouar2015}%
  \BibitemOpen
  \bibfield  {author} {\bibinfo {author} {\bibfnamefont {S.}~\bibnamefont {Bounouar}}, \bibinfo {author} {\bibfnamefont {M.}~\bibnamefont {M\"uller}}, \bibinfo {author} {\bibfnamefont {A.~M.}\ \bibnamefont {Barth}}, \bibinfo {author} {\bibfnamefont {M.}~\bibnamefont {Gl\"assl}}, \bibinfo {author} {\bibfnamefont {V.~M.}\ \bibnamefont {Axt}}, \ and\ \bibinfo {author} {\bibfnamefont {P.}~\bibnamefont {Michler}},\ }\bibfield  {title} {\enquote {\bibinfo {title} {Phonon-assisted robust and deterministic two-photon biexciton preparation in a quantum dot},}\ }\href {\doibase 10.1103/PhysRevB.91.161302} {\bibfield  {journal} {\bibinfo  {journal} {Phys. Rev. B}\ }\textbf {\bibinfo {volume} {91}},\ \bibinfo {pages} {161302} (\bibinfo {year} {2015})}\BibitemShut {NoStop}%
\bibitem [{\citenamefont {Hopfmann}\ \emph {et~al.}(2021{\natexlab{b}})\citenamefont {Hopfmann}, \citenamefont {Sharma}, \citenamefont {Nie}, \citenamefont {Keil}, \citenamefont {Ding},\ and\ \citenamefont {Schmidt}}]{Hopfmann2021a}%
  \BibitemOpen
  \bibfield  {author} {\bibinfo {author} {\bibfnamefont {C.}~\bibnamefont {Hopfmann}}, \bibinfo {author} {\bibfnamefont {N.~L.}\ \bibnamefont {Sharma}}, \bibinfo {author} {\bibfnamefont {W.}~\bibnamefont {Nie}}, \bibinfo {author} {\bibfnamefont {R.}~\bibnamefont {Keil}}, \bibinfo {author} {\bibfnamefont {F.}~\bibnamefont {Ding}}, \ and\ \bibinfo {author} {\bibfnamefont {O.~G.}\ \bibnamefont {Schmidt}},\ }\bibfield  {title} {\enquote {\bibinfo {title} {Heralded preparation of spin qubits in droplet-etched {GaAs} quantum dots using quasiresonant excitation},}\ }\href {\doibase 10.1103/PhysRevB.104.075301} {\bibfield  {journal} {\bibinfo  {journal} {Phys. Rev. B}\ }\textbf {\bibinfo {volume} {104}},\ \bibinfo {pages} {075301} (\bibinfo {year} {2021}{\natexlab{b}})}\BibitemShut {NoStop}%
\bibitem [{\citenamefont {Langer}\ \emph {et~al.}(2025)\citenamefont {Langer}, \citenamefont {Dhurjati}, \citenamefont {Zena}, \citenamefont {Rahimi}, \citenamefont {Pal}, \citenamefont {Raith}, \citenamefont {Nestler}, \citenamefont {Bassoli}, \citenamefont {Fitzek}, \citenamefont {Schmidt},\ and\ \citenamefont {Hopfmann}}]{Langer2025a}%
  \BibitemOpen
  \bibfield  {author} {\bibinfo {author} {\bibfnamefont {M.}~\bibnamefont {Langer}}, \bibinfo {author} {\bibfnamefont {S.~A.}\ \bibnamefont {Dhurjati}}, \bibinfo {author} {\bibfnamefont {Y.~G.}\ \bibnamefont {Zena}}, \bibinfo {author} {\bibfnamefont {A.}~\bibnamefont {Rahimi}}, \bibinfo {author} {\bibfnamefont {M.}~\bibnamefont {Pal}}, \bibinfo {author} {\bibfnamefont {L.}~\bibnamefont {Raith}}, \bibinfo {author} {\bibfnamefont {S.}~\bibnamefont {Nestler}}, \bibinfo {author} {\bibfnamefont {R.}~\bibnamefont {Bassoli}}, \bibinfo {author} {\bibfnamefont {F.~H.~P.}\ \bibnamefont {Fitzek}}, \bibinfo {author} {\bibfnamefont {O.~G.}\ \bibnamefont {Schmidt}}, \ and\ \bibinfo {author} {\bibfnamefont {C.}~\bibnamefont {Hopfmann}},\ }\href {https://arxiv.org/abs/2503.07305} {\enquote {\bibinfo {title} {Bright quantum dot light sources using monolithic microlenses on gold back-reflectors},}\ } (\bibinfo {year} {2025}),\ \Eprint {http://arxiv.org/abs/2503.07305} {arXiv:2503.07305} \BibitemShut {NoStop}%
\bibitem [{\citenamefont {Gissibl}\ \emph {et~al.}(2016)\citenamefont {Gissibl}, \citenamefont {Thiele}, \citenamefont {Herkommer},\ and\ \citenamefont {Giessen}}]{Gissibl2016}%
  \BibitemOpen
  \bibfield  {author} {\bibinfo {author} {\bibfnamefont {T.}~\bibnamefont {Gissibl}}, \bibinfo {author} {\bibfnamefont {S.}~\bibnamefont {Thiele}}, \bibinfo {author} {\bibfnamefont {A.}~\bibnamefont {Herkommer}}, \ and\ \bibinfo {author} {\bibfnamefont {H.}~\bibnamefont {Giessen}},\ }\bibfield  {title} {\enquote {\bibinfo {title} {Sub-micrometre accurate free-form optics by three-dimensional printing on single-mode fibres},}\ }\href {\doibase 10.1038/ncomms11763} {\bibfield  {journal} {\bibinfo  {journal} {Nature Communications}\ }\textbf {\bibinfo {volume} {7}},\ \bibinfo {pages} {11763} (\bibinfo {year} {2016})}\BibitemShut {NoStop}%
\bibitem [{\citenamefont {Sartison}\ \emph {et~al.}(2021)\citenamefont {Sartison}, \citenamefont {Weber}, \citenamefont {Thiele}, \citenamefont {Bremer}, \citenamefont {Fischbach}, \citenamefont {Herzog}, \citenamefont {Kolatschek}, \citenamefont {Jetter}, \citenamefont {Reitzenstein}, \citenamefont {Herkommer}, \citenamefont {Michler}, \citenamefont {Portalupi},\ and\ \citenamefont {Giessen}}]{Sartison2021}%
  \BibitemOpen
  \bibfield  {author} {\bibinfo {author} {\bibfnamefont {M.}~\bibnamefont {Sartison}}, \bibinfo {author} {\bibfnamefont {K.}~\bibnamefont {Weber}}, \bibinfo {author} {\bibfnamefont {S.}~\bibnamefont {Thiele}}, \bibinfo {author} {\bibfnamefont {L.}~\bibnamefont {Bremer}}, \bibinfo {author} {\bibfnamefont {S.}~\bibnamefont {Fischbach}}, \bibinfo {author} {\bibfnamefont {T.}~\bibnamefont {Herzog}}, \bibinfo {author} {\bibfnamefont {S.}~\bibnamefont {Kolatschek}}, \bibinfo {author} {\bibfnamefont {M.}~\bibnamefont {Jetter}}, \bibinfo {author} {\bibfnamefont {S.}~\bibnamefont {Reitzenstein}}, \bibinfo {author} {\bibfnamefont {A.}~\bibnamefont {Herkommer}}, \bibinfo {author} {\bibfnamefont {P.}~\bibnamefont {Michler}}, \bibinfo {author} {\bibfnamefont {S.~L.}\ \bibnamefont {Portalupi}}, \ and\ \bibinfo {author} {\bibfnamefont {H.}~\bibnamefont {Giessen}},\ }\bibfield  {title} {\enquote {\bibinfo {title} {{3D printed micro-optics for quantum technology: Optimised coupling of single quantum dot emission into a
  single-mode fibre}},}\ }\href {\doibase 10.37188/lam.2021.006} {\bibfield  {journal} {\bibinfo  {journal} {Light: Advanced Manufacturing}\ }\textbf {\bibinfo {volume} {2}},\ \bibinfo {pages} {103} (\bibinfo {year} {2021})}\BibitemShut {NoStop}%
\bibitem [{\citenamefont {Weber}\ \emph {et~al.}(2024)\citenamefont {Weber}, \citenamefont {Thiele}, \citenamefont {Hentschel}, \citenamefont {Herkommer},\ and\ \citenamefont {Giessen}}]{Weber2024}%
  \BibitemOpen
  \bibfield  {author} {\bibinfo {author} {\bibfnamefont {K.}~\bibnamefont {Weber}}, \bibinfo {author} {\bibfnamefont {S.}~\bibnamefont {Thiele}}, \bibinfo {author} {\bibfnamefont {M.}~\bibnamefont {Hentschel}}, \bibinfo {author} {\bibfnamefont {A.}~\bibnamefont {Herkommer}}, \ and\ \bibinfo {author} {\bibfnamefont {H.}~\bibnamefont {Giessen}},\ }\bibfield  {title} {\enquote {\bibinfo {title} {Positional accuracy of {3D} printed quantum emitter fiber couplers},}\ }\href {\doibase https://doi.org/10.1002/qute.202400135} {\bibfield  {journal} {\bibinfo  {journal} {Advanced Quantum Technologies}\ }\textbf {\bibinfo {volume} {n/a}},\ \bibinfo {pages} {2400135} (\bibinfo {year} {2024})}\BibitemShut {NoStop}%
\bibitem [{\citenamefont {Schwab}\ \emph {et~al.}(2022)\citenamefont {Schwab}, \citenamefont {Weber}, \citenamefont {Drozella}, \citenamefont {Jimenez}, \citenamefont {Herkommer}, \citenamefont {Bremer}, \citenamefont {Reitzenstein},\ and\ \citenamefont {Giessen}}]{Schwab2022}%
  \BibitemOpen
  \bibfield  {author} {\bibinfo {author} {\bibfnamefont {J.}~\bibnamefont {Schwab}}, \bibinfo {author} {\bibfnamefont {K.}~\bibnamefont {Weber}}, \bibinfo {author} {\bibfnamefont {J.}~\bibnamefont {Drozella}}, \bibinfo {author} {\bibfnamefont {C.}~\bibnamefont {Jimenez}}, \bibinfo {author} {\bibfnamefont {A.}~\bibnamefont {Herkommer}}, \bibinfo {author} {\bibfnamefont {L.}~\bibnamefont {Bremer}}, \bibinfo {author} {\bibfnamefont {S.}~\bibnamefont {Reitzenstein}}, \ and\ \bibinfo {author} {\bibfnamefont {H.}~\bibnamefont {Giessen}},\ }\bibfield  {title} {\enquote {\bibinfo {title} {Coupling light emission of single-photon sources into single-mode fibers: mode matching, coupling efficiencies, and thermo-optical effects},}\ }\href {\doibase 10.1364/OE.465101} {\bibfield  {journal} {\bibinfo  {journal} {Opt. Express}\ }\textbf {\bibinfo {volume} {30}},\ \bibinfo {pages} {32292--32305} (\bibinfo {year} {2022})}\BibitemShut {NoStop}%
\bibitem [{\citenamefont {Asadollahbaik}\ \emph {et~al.}(2020)\citenamefont {Asadollahbaik}, \citenamefont {Thiele}, \citenamefont {Weber}, \citenamefont {Kumar}, \citenamefont {Drozella}, \citenamefont {Sterl}, \citenamefont {Herkommer}, \citenamefont {Giessen},\ and\ \citenamefont {Fick}}]{Asadollahbaik2020}%
  \BibitemOpen
  \bibfield  {author} {\bibinfo {author} {\bibfnamefont {A.}~\bibnamefont {Asadollahbaik}}, \bibinfo {author} {\bibfnamefont {S.}~\bibnamefont {Thiele}}, \bibinfo {author} {\bibfnamefont {K.}~\bibnamefont {Weber}}, \bibinfo {author} {\bibfnamefont {A.}~\bibnamefont {Kumar}}, \bibinfo {author} {\bibfnamefont {J.}~\bibnamefont {Drozella}}, \bibinfo {author} {\bibfnamefont {F.}~\bibnamefont {Sterl}}, \bibinfo {author} {\bibfnamefont {A.~M.}\ \bibnamefont {Herkommer}}, \bibinfo {author} {\bibfnamefont {H.}~\bibnamefont {Giessen}}, \ and\ \bibinfo {author} {\bibfnamefont {J.}~\bibnamefont {Fick}},\ }\bibfield  {title} {\enquote {\bibinfo {title} {Highly efficient dual-fiber optical trapping with {3D} printed diffractive fresnel lenses},}\ }\href {\doibase 10.1021/acsphotonics.9b01024} {\bibfield  {journal} {\bibinfo  {journal} {ACS Photonics}\ }\textbf {\bibinfo {volume} {7}},\ \bibinfo {pages} {88--97} (\bibinfo {year} {2020})}\BibitemShut {NoStop}%
\bibitem [{\citenamefont {Ruchka}\ \emph {et~al.}(2022)\citenamefont {Ruchka}, \citenamefont {Hammer}, \citenamefont {Rockenhäuser}, \citenamefont {Albrecht}, \citenamefont {Drozella}, \citenamefont {Thiele}, \citenamefont {Giessen},\ and\ \citenamefont {Langen}}]{Ruchka2022}%
  \BibitemOpen
  \bibfield  {author} {\bibinfo {author} {\bibfnamefont {P.}~\bibnamefont {Ruchka}}, \bibinfo {author} {\bibfnamefont {S.}~\bibnamefont {Hammer}}, \bibinfo {author} {\bibfnamefont {M.}~\bibnamefont {Rockenhäuser}}, \bibinfo {author} {\bibfnamefont {R.}~\bibnamefont {Albrecht}}, \bibinfo {author} {\bibfnamefont {J.}~\bibnamefont {Drozella}}, \bibinfo {author} {\bibfnamefont {S.}~\bibnamefont {Thiele}}, \bibinfo {author} {\bibfnamefont {H.}~\bibnamefont {Giessen}}, \ and\ \bibinfo {author} {\bibfnamefont {T.}~\bibnamefont {Langen}},\ }\bibfield  {title} {\enquote {\bibinfo {title} {Microscopic {3D} printed optical tweezers for atomic quantum technology},}\ }\href {\doibase 10.1088/2058-9565/ac796c} {\bibfield  {journal} {\bibinfo  {journal} {Quantum Science and Technology}\ }\textbf {\bibinfo {volume} {7}},\ \bibinfo {pages} {045011} (\bibinfo {year} {2022})}\BibitemShut {NoStop}%
\bibitem [{Note1()}]{Note1}%
  \BibitemOpen
  \bibinfo {note} {Fiber diameter \qty {250}{\micro \metre } with polyamide coating of and \qty {125}{\micro \metre } bare.}\BibitemShut {Stop}%
\bibitem [{\citenamefont {Winik}\ \emph {et~al.}(2017)\citenamefont {Winik}, \citenamefont {Cogan}, \citenamefont {Don}, \citenamefont {Schwartz}, \citenamefont {Gantz}, \citenamefont {Schmidgall}, \citenamefont {Livneh}, \citenamefont {Rapaport}, \citenamefont {Buks},\ and\ \citenamefont {Gershoni}}]{Winik2017}%
  \BibitemOpen
  \bibfield  {author} {\bibinfo {author} {\bibfnamefont {R.}~\bibnamefont {Winik}}, \bibinfo {author} {\bibfnamefont {D.}~\bibnamefont {Cogan}}, \bibinfo {author} {\bibfnamefont {Y.}~\bibnamefont {Don}}, \bibinfo {author} {\bibfnamefont {I.}~\bibnamefont {Schwartz}}, \bibinfo {author} {\bibfnamefont {L.}~\bibnamefont {Gantz}}, \bibinfo {author} {\bibfnamefont {E.~R.}\ \bibnamefont {Schmidgall}}, \bibinfo {author} {\bibfnamefont {N.}~\bibnamefont {Livneh}}, \bibinfo {author} {\bibfnamefont {R.}~\bibnamefont {Rapaport}}, \bibinfo {author} {\bibfnamefont {E.}~\bibnamefont {Buks}}, \ and\ \bibinfo {author} {\bibfnamefont {D.}~\bibnamefont {Gershoni}},\ }\bibfield  {title} {\enquote {\bibinfo {title} {On-demand source of maximally entangled photon pairs using the biexciton-exciton radiative cascade},}\ }\href {\doibase 10.1103/PhysRevB.95.235435} {\bibfield  {journal} {\bibinfo  {journal} {Phys. Rev. B}\ }\textbf {\bibinfo {volume} {95}},\ \bibinfo {pages} {235435} (\bibinfo {year} {2017})}\BibitemShut {NoStop}%
\bibitem [{Note2()}]{Note2}%
  \BibitemOpen
  \bibinfo {note} {For definition of the 6 polarization bases H,V,D,A,R and L see \protect \cref {sec:prec_osc}.}\BibitemShut {Stop}%
\bibitem [{\citenamefont {James}\ \emph {et~al.}(2001)\citenamefont {James}, \citenamefont {Kwiat}, \citenamefont {Munro},\ and\ \citenamefont {White}}]{James2001}%
  \BibitemOpen
  \bibfield  {author} {\bibinfo {author} {\bibfnamefont {D.~F.~V.}\ \bibnamefont {James}}, \bibinfo {author} {\bibfnamefont {P.~G.}\ \bibnamefont {Kwiat}}, \bibinfo {author} {\bibfnamefont {W.~J.}\ \bibnamefont {Munro}}, \ and\ \bibinfo {author} {\bibfnamefont {A.~G.}\ \bibnamefont {White}},\ }\bibfield  {title} {\enquote {\bibinfo {title} {Measurement of qubits},}\ }\href {\doibase 10.1103/PhysRevA.64.052312} {\bibfield  {journal} {\bibinfo  {journal} {Phys. Rev. A}\ }\textbf {\bibinfo {volume} {64}},\ \bibinfo {pages} {052312} (\bibinfo {year} {2001})}\BibitemShut {NoStop}%
\bibitem [{\citenamefont {Keil}\ \emph {et~al.}(2017)\citenamefont {Keil}, \citenamefont {Zopf}, \citenamefont {Chen}, \citenamefont {H{\"o}fer}, \citenamefont {Zhang}, \citenamefont {Ding},\ and\ \citenamefont {Schmidt}}]{Keil2017}%
  \BibitemOpen
  \bibfield  {author} {\bibinfo {author} {\bibfnamefont {R.}~\bibnamefont {Keil}}, \bibinfo {author} {\bibfnamefont {M.}~\bibnamefont {Zopf}}, \bibinfo {author} {\bibfnamefont {Y.}~\bibnamefont {Chen}}, \bibinfo {author} {\bibfnamefont {B.}~\bibnamefont {H{\"o}fer}}, \bibinfo {author} {\bibfnamefont {J.}~\bibnamefont {Zhang}}, \bibinfo {author} {\bibfnamefont {F.}~\bibnamefont {Ding}}, \ and\ \bibinfo {author} {\bibfnamefont {O.~G.}\ \bibnamefont {Schmidt}},\ }\bibfield  {title} {\enquote {\bibinfo {title} {Solid-state ensemble of highly entangled photon sources at rubidium atomic transitions},}\ }\href@noop {} {\bibfield  {journal} {\bibinfo  {journal} {Nat. Commun.}\ }\textbf {\bibinfo {volume} {8}},\ \bibinfo {pages} {1--8} (\bibinfo {year} {2017})}\BibitemShut {NoStop}%
\bibitem [{\citenamefont {Gschrey}\ \emph {et~al.}(2015)\citenamefont {Gschrey}, \citenamefont {Thoma}, \citenamefont {Schnauber}, \citenamefont {Seifried}, \citenamefont {Schmidt}, \citenamefont {Wohlfeil}, \citenamefont {Krüger}, \citenamefont {Schulze}, \citenamefont {Heindel}, \citenamefont {Burger}, \citenamefont {Schmidt}, \citenamefont {Strittmatter}, \citenamefont {Rodt},\ and\ \citenamefont {Reitzenstein}}]{Gschrey2015}%
  \BibitemOpen
  \bibfield  {author} {\bibinfo {author} {\bibfnamefont {M.}~\bibnamefont {Gschrey}}, \bibinfo {author} {\bibfnamefont {A.}~\bibnamefont {Thoma}}, \bibinfo {author} {\bibfnamefont {P.}~\bibnamefont {Schnauber}}, \bibinfo {author} {\bibfnamefont {M.}~\bibnamefont {Seifried}}, \bibinfo {author} {\bibfnamefont {R.}~\bibnamefont {Schmidt}}, \bibinfo {author} {\bibfnamefont {B.}~\bibnamefont {Wohlfeil}}, \bibinfo {author} {\bibfnamefont {L.}~\bibnamefont {Krüger}}, \bibinfo {author} {\bibfnamefont {J.~H.}\ \bibnamefont {Schulze}}, \bibinfo {author} {\bibfnamefont {T.}~\bibnamefont {Heindel}}, \bibinfo {author} {\bibfnamefont {S.}~\bibnamefont {Burger}}, \bibinfo {author} {\bibfnamefont {F.}~\bibnamefont {Schmidt}}, \bibinfo {author} {\bibfnamefont {A.}~\bibnamefont {Strittmatter}}, \bibinfo {author} {\bibfnamefont {S.}~\bibnamefont {Rodt}}, \ and\ \bibinfo {author} {\bibfnamefont {S.}~\bibnamefont {Reitzenstein}},\ }\bibfield  {title} {\enquote {\bibinfo {title} {Highly indistinguishable photons from
  deterministic quantum-dot microlenses utilizing three-dimensional in situ electron-beam lithography},}\ }\href {\doibase 10.1038/ncomms8662} {\bibfield  {journal} {\bibinfo  {journal} {Nature Communications}\ }\textbf {\bibinfo {volume} {6}},\ \bibinfo {pages} {7662} (\bibinfo {year} {2015})}\BibitemShut {NoStop}%
\bibitem [{\citenamefont {Bayer}\ \emph {et~al.}(2002)\citenamefont {Bayer}, \citenamefont {Ortner}, \citenamefont {Stern}, \citenamefont {Kuther}, \citenamefont {Gorbunov}, \citenamefont {Forchel}, \citenamefont {Hawrylak}, \citenamefont {Fafard}, \citenamefont {Hinzer}, \citenamefont {Reinecke}, \citenamefont {Walck}, \citenamefont {Reithmaier}, \citenamefont {Klopf},\ and\ \citenamefont {Sch\"afer}}]{Bayer2002}%
  \BibitemOpen
  \bibfield  {author} {\bibinfo {author} {\bibfnamefont {M.}~\bibnamefont {Bayer}}, \bibinfo {author} {\bibfnamefont {G.}~\bibnamefont {Ortner}}, \bibinfo {author} {\bibfnamefont {O.}~\bibnamefont {Stern}}, \bibinfo {author} {\bibfnamefont {A.}~\bibnamefont {Kuther}}, \bibinfo {author} {\bibfnamefont {A.~A.}\ \bibnamefont {Gorbunov}}, \bibinfo {author} {\bibfnamefont {A.}~\bibnamefont {Forchel}}, \bibinfo {author} {\bibfnamefont {P.}~\bibnamefont {Hawrylak}}, \bibinfo {author} {\bibfnamefont {S.}~\bibnamefont {Fafard}}, \bibinfo {author} {\bibfnamefont {K.}~\bibnamefont {Hinzer}}, \bibinfo {author} {\bibfnamefont {T.~L.}\ \bibnamefont {Reinecke}}, \bibinfo {author} {\bibfnamefont {S.~N.}\ \bibnamefont {Walck}}, \bibinfo {author} {\bibfnamefont {J.~P.}\ \bibnamefont {Reithmaier}}, \bibinfo {author} {\bibfnamefont {F.}~\bibnamefont {Klopf}}, \ and\ \bibinfo {author} {\bibfnamefont {F.}~\bibnamefont {Sch\"afer}},\ }\bibfield  {title} {\enquote {\bibinfo {title} {{Fine structure of neutral and charged excitons in
  self-assembled In(Ga)As/(Al)GaAs quantum dots}},}\ }\href {\doibase 10.1103/PhysRevB.65.195315} {\bibfield  {journal} {\bibinfo  {journal} {Phys. Rev. B}\ }\textbf {\bibinfo {volume} {65}},\ \bibinfo {pages} {195315} (\bibinfo {year} {2002})}\BibitemShut {NoStop}%
\end{thebibliography}%

\end{document}